\UseRawInputEncoding
\documentclass[showpacs,twocolumn,superscriptaddress]{revtex4} 

\usepackage{doi}
\usepackage{hyperref}
\hypersetup{colorlinks=true,linkcolor=blue,citecolor=cyan}
\usepackage{graphicx}
\usepackage{xcolor,soul}
\sethlcolor{green}
\usepackage{dcolumn}
\usepackage{bm}
\usepackage{color}
\usepackage{enumitem}
\usepackage{mathrsfs}
\usepackage{amsmath}
\usepackage{cleveref}
\usepackage{amssymb}
\usepackage{appendix}

\crefname{equation}{Eqn.}{Eqns.}
\crefname{figure}{Fig.}{Figs.}
\crefname{section}{Sec.}{Sec.}
\crefname{table}{Table}{Tables}
\crefname{appsec}{Appendix}{Appendices}
\usepackage{physics}
\usepackage{cancel}

\setstcolor{red}

\begin{document}

\title{ Motion of charged and spinning particles influenced by dark matter field surrounding a charged 
dyonic black hole
}
\author{Sanjar Shaymatov}
\email{sanjar@astrin.uz}
\affiliation{Institute for Theoretical Physics and Cosmology, Zheijiang University of Technology, Hangzhou 310023, China}
\affiliation{Akfa University,  Milliy Bog Street 264, Tashkent 111221, Uzbekistan}
\affiliation{Ulugh Beg Astronomical Institute, Astronomicheskaya 33, Tashkent 100052, Uzbekistan} 
\affiliation{National University of Uzbekistan, Tashkent 100174, Uzbekistan} 
\affiliation{Institute of Fundamental and Applied Research, National Research University TIIAME,
Kori Niyoziy 39, Tashkent 100000, Uzbekistan}
\author{Pankaj Sheoran}
\email{hukmipankaj@gmail.com}
\affiliation{Department of Physics, Institute of Science, Banaras Hindu University, Varanasi-221005, India}
\author{Sanjay Siwach}
\email{sksiwach@hotmail.com}
\affiliation{Department of Physics, Institute of Science, Banaras Hindu University, Varanasi-221005, India}

\date{\today}
\begin{abstract} 
We investigate the motion of massive charged and spinning test particles around a charged dyonic black hole 
surrounded by perfect fluid scalar dark matter field. We obtain the equations of motion and find the expressions for the four-velocity for the case of a charged particle and four-momentum components for the case of a spinning particle. The trajectories for various values of electric $Q_{e}$ and magnetic $Q_{m}$ charges are investigated under the influence of dark matter field $\lambda$. 
We study in detail the properties of innermost stable circular orbits (ISCOs) in the equatorial plane. 
We show that, in addition to the particle's spin, the dark matter field parameter $\lambda$ and black hole charges ($Q_{m}\;\text{and}\;Q_{e}$) have a significant influence on the ISCOs of spinning particles. 
{We find} that if the spin is parallel to the total angular momentum $J$ (i.e., $\mathcal{S}>0$), the ISCO parameters (i.e., $r_{ISCO}, L_{ISCO},\;\text{and}\; E_{ISCO}$) of a spinning particle are smaller than those of a nonspinning particle, whereas if the spin is antiparallel to total angular momentum $J$ (i.e. $\mathcal{S}<0$), the value of the ISCO parameters is greater than that of the nonspinning particle. We also show that for the corresponding values of spin parameter S, the behavior of Keplerian angular frequency {$\Omega$ at} the ISCO  is opposite that of  $r_{ISCO}, L_{ISCO},\; \text{and}\; E_{ISCO}$. As an astrophysical application, the study of ISCOs may play a significant role as ISCOs can be used to set the inner edge of the accretion disk around a black hole as well as the initial condition of the binary black hole mergers.
\end{abstract}


\maketitle

\section{Introduction \label{Sec:introduction}}

Black holes have so far been predicted to be formed under gravitational collapse of a sufficiently compact massive object that has exerted all its resources to withstand gravity, and thus they have been very attractive objects with their remarkable geometric properties. Recent observational studies of stellar mass black holes in x-ray binaries and gravitational wave astronomy \cite{Abbott16a,Abbott16b} and supermassive black holes through modern astronomical observations of the Event Horizon Telescope Collaboration and BlackHoleCam \cite{Akiyama19L1,Akiyama19L6} have verified the presence of black holes in the Universe.  Since much progress has been made in observations of the first image of a supermassive black hole at the center of the M87$^{\star}$ galaxy, black holes have been currently considered the main laboratories in testing general relativity and modified theories of gravity. Because of these abovementioned observations, a qualitatively new stage has been opened to study the remarkable gravitational properties of astrophysical black holes and obtain precise constraints and measurement of black hole parameters. 

To understand more deeply and explore the properties of black holes, one needs to investigate the {geodesics} 
of test particles and photons as their motion would be a very potent tool in addressing a question in connection with the black hole's nature.  In this respect, the {test} particle {trajectories} 
may play a decisive role in understanding the unexplored nature of other existing fields contributing to altering its motion as well as the behavior of background 
{spacetime} in the astrophysical context. For example, the motion of test particles gets altered drastically as a consequence of the presence of external magnetic fields \cite{Benavides-Gallego19} and the presence of surrounding matter fields
\cite{Tsukamoto13,Bini12}.
On the other hand, particle motion around black holes gives the possibility of exhibiting departures of the geometry of astrophysical black holes.  There is an extensive body of research involving the effect of the electromagnetic field on the motion of a test particle in the black hole vicinity in various gravity models (, for example
\cite{Kovar08,Shaymatov15,Dadhich18,Pavlovic19,Shaymatov19b,Shaymatov20egb,Turimov2019PRD,Rayimbaev20c}).
Furthermore, one of the other existing matters is the dark matter field that can exist in the environment of supermassive black holes. 
With this in mind, one is allowed to test the effect of the dark matter field on both the background geometry as well as the test particle motion in a realistic astrophysical context. 
Regardless of the fact that there is still no direct experimental detection that verifies the presence of dark matter, observations have provided strong evidence that dark matter can exist in the environment of giant elliptical and spiral galaxies~\cite{Rubin80}. Relying on the theoretical analysis and astrophysical data, it is believed that dark matter contributes approximately up to 90 \% of the mass of the Galaxy, while the rest is the luminous matter composed of baryonic matter~\cite{Persic96}. Astrophysical observations indicate that giant elliptical and spiral galaxies are embedded in a giant dark matter halo~\cite{Akiyama19L1,Akiyama19L6}. 
In the literature, there are several black hole solutions that involve the dark matter contribution in the background geometry; here we give the representative ones~\cite{Kiselev03,Li-Yang12}. 
Of them, Kiselev derived a static and spherically symmetric black hole with a dark matter profile through a quintessential scalar field~\cite{Kiselev03}, according to which the new solution involved a term $\lambda\ln(r/r_q)$ in the metric coefficients. This solution, in turn gave new prosperity in deriving further solutions. Later on, another interesting solution was derived by Li and Yang~\cite{Li-Yang12} containing the same term $(r_q/r)\ln(r/r_q)$ represented by a phantom scalar field under the assumption that the dark matter profile is formed from massive particles that must participate in the weak interaction (i.e. weakly interacting massive particles with the equation of state $\omega\simeq 0$). There is a large amount of work the considers the dark matter field in the background {spacetime} in various situations ~(see,
e.g.\;\cite{Xu18,Haroon19,Konoplya19plb,Hendi20,Jusufi19,Shaymatov21d,Rayimbaev-Shaymatov21a,Shaymatov21pdu}).

In an astrophysical scenario, black holes can be characterized by at most three parameters, i.e. black hole mass $M$, rotation $a$ and electric charge $Q$. Note that the first two parameters are constrained by the black hole mass. 
Also black holes can be regarded as charged black holes referred to as the Reissner-Nordstr\"{o}m black hole solution with mass and electric charge. 
For example, the detailed analysis of the neutral and charged particle motion around a Reissner-Nordstr\"{o}m black hole has been discussed in~\cite{Pugliese11,Pugliese11b}. There are several astrophysical mechanisms for a black hole to possess charge. A black hole can have a positive net electric charge due to the balance between the Coulomb and gravitational forces for charged particles near the surface of the compact object~ \cite{Zajacek19,Bally78} and due to the matter which gets charged as that of the irradiating photons~\cite{Weingartner06}. It is also worth noting that, by the Wald mechanism \cite{Wald74} the induced charge can be produced by the magnetic field lines getting twisted as that of the fame-dragging effect. However, in all mentioned cases the black hole charge can be regarded as much more weaker (see, for example, \cite{Zajacek18}). Also, based on the an exact rotating Schwinger dyon solution, a black hole can be endowed with four parameters: mass $M$, the electric charge $Q_e$, magnetic charge $Q_m$, and rotation parameter $a$~\cite{Kasuya82}. This black hole solution is an interesting case of a charged black hole even if there exists no rotation parameter. In this context, there are several other ways that a black hole can include electric and magnetic charges, and the properties of these black hole solutions have been tested by different astrophysical processes 
\cite{Narzilloev20a,Narzilloev20b}.

The analysis of particle motion is a powerful tool to reveal the geometric properties of compact objects and astrophysical processes in the environment surrounding the black holes \cite{Frolov10,Aliev02,Shaymatov14,Duztas-Jamil20,Shaymatov19c,Shaymatov21b}.
In a realistic scenario background spacetime geometry cannot be regarded as vacuum. Thus, in this paper, we consider a charged dyonic black hole placed in the perfect fluid dark matter (PFDM) field. For this spacetime geometry, we study the influence of PFDM on the dynamical motion of magnetically and electrically charged particles moving around this charged dyonic black hole. This may lead us to understand the nature of the gravitational interaction between the charged black hole spacetime and PFDM field. In addition to the motion of a charged particle, we also investigate the motion of a spinning particle moving in the vicinity of a charged dyonic black hole surrounded by a PFDM. For spinning particles, we mainly discuss the properties of innermost stable circular orbit (ISCO) parameters $r_{ISCO}, L_{ISCO}, E_{ISCO},\;\text{and}\; \Omega_{ISCO}$ and bring out the effect of dark matter field in addition to the black hole charges. 
The main motivation to study the ISCOs is that they can be treated as the initial condition of the final merger of a binary system of compact objects.
Obviously, the motion of a massive test particle is influenced by black hole parameters like charge, mass and spin (see Refs. \cite{Suzuki:1997by,Campanelli:2006gf,Cabanac:2009yz,Akcay:2012ea,Chakraborty:2013kza,Zaslavskii:2014mqa,Delsate:2015ina,Shaymatov21c,Dadhich22a}).
The study of the ISCOs of the spinning test particle started with the works of Suzuki and Maeda \cite{Suzuki:1997by} and Jefremove \textit{et al.} \cite{Jefremov:2015gza}. In \cite{Suzuki:1997by}, the ISCOs of a spinning test particle are explored for Kerr a black hole. Later, in \cite{Jefremov:2015gza}, the approximate analytical solutions of the ISCOs for a spinning particle are presented for both Schwarzschild and Kerr black holes within the small spin limit. Thereafter, the study of ISCOs for spinning particles in various black hole backgrounds is done
\cite{dAmbrosi:2015wqz,Harms:2016ctx,Lukes-Gerakopoulos:2017vkj,Zhang:2017nhl,Toshmatov:2019bda,Nucamendi:2019qsn,Conde:2019juj,Larranaga:2020ycg,Zhang:2020qew,Zhang:2020hzu},
although, the motion of a spinning particle is studied in various contexts
 \cite{Suzuki:1999si,Singh:2005ha,Koyama:2007cj,Damour:2008qf,Singh:2008qs,Singh:2008qr,Harms:2015ixa,Mukherjee:2018zug,Okabayashi:2019wjs,Sheoran:2020kmn,Pankaj:2021,Shahzadi:2021upd}.
Still, there are not many studies of the ISCOs of a spinning particle when a black hole with the surrounding medium is considered.

{Here}, we are interested in to seeing how the surrounding medium (say, PFDM in our case) will affect the motion of a spinning particle besides the black hole parameters. It is well known that when the test particle reaction effects (such as self-force corrections) are taken into account, the test particle will follow a nongeodesic trajectory \cite{Warburton:2011hp,Isoyama:2014mja,vandeMeent:2016hel}. Additionally, if the test particle has a spin due to spin-curvature coupling an extra force is acting on the particle which also results in nongeodesic motion of the spinning test particle \cite{Hanson:1974qy,Wald:1972sz}. 

In this paper, we consider only the spin-orbit coupling via ``pole-dipole" approximation and discard the reaction of the particle with the black hole background in order to numerically investigate the properties the ISCO of a spinning particle with an arbitrary value of the spin $S$ moving in the charged dyonic black hole immersed in the PFDM.  
To do so, we use the Tulczyjew spin-supplementary condition (TSSC). We investigate the effect of $\lambda$, $Q_{m}$, $Q_{e}$, and $S$ on the ISCO parameters  $r_{ISCO}, L_{ISCO}, E_{ISCO},\;\text{and}\; \Omega_{ISCO}$ of a spinning test particle together with the superluminal constraint (for a brief discussion on the equations of motion of the spinning particle in curved spacetime, TSSC and superluminal constraint, see Appendix \ref{Sec:Appendix_A} and references therein). It is worth noting that here we are considering the spin orbit coupling only up to first order and discard the higher order spin correction terms known as the $1/c^{2}$-approximation (or as quadratic spin corrections) \cite{Deriglazov:2017jub,Deriglazov:2018vwa}.

The paper is organized as follows: In Sec.~\ref{Sec:metric}, we describe briefly the charged black hole metric which is followed by the study of charged particle dynamics in Sec.~\ref{Sec:motion}.  
We investigate the motion of a spinning particle around a charged black hole submerged in a perfect fluid dark matter environment in \cref{sec:spinning_particles}. In \cref{Sec:conclusion}, we offer a summary and emphasize the important findings. To make the article self-contained, we include a brief introduction of the Lagrangian theory of the spinning particle in \cref{Sec:Appendix_A}. In addition, the explicit form of the equations required to identify the ISCOs and the explicit form of the equation expressing the superluminal constraint are provided in \cref{Sec:Appendix_B}. Throughout this work, we use a system of units in which $G=c=1$. Greek indices are taken to run from 0 to 3, while latin ones from 1 to 3.

\section{The black hole metric\label{Sec:metric}}	
We consider the Lagrangian density of Einstein-Maxwell gravity in the presence of the perfect fluid scalar dark matter field \cite{Kamata81,Li-Yang12}
\begin{align}\nonumber
\mathscr{S}&=\frac{1}{16\pi}\int d^4x\sqrt{-g}\Big[R-F_{\mu\nu}F^{\mu\nu}\\&-2\left(\nabla_\mu\Phi\nabla^\mu\Phi-2V(\Phi)\right)-4(\mathscr{L}_{DM}+\mathscr{L}_{I})\Big]\, , \label{eq:action} 
\end{align}
with the electromagnetic field tensor
\begin{align}\label{eq:N2}
F_{~\mu\nu}=\partial_{\mu}A_{\nu}-\partial_{\nu}A_{\mu}+^{\ast}G_{\mu\nu}\, , 
\end{align}
where $^{\ast}G_{\mu\nu}$ refers to the Dirac string term \cite{Kamata81}. Note that $V(\Phi)$ is related to the phantom field potential, while  $\mathcal{L}_{\rm DM}$ and $\mathscr{L}_{I}$, respectively, refer to the dark matter Lagrangian density and Lagrangian representing the interaction between the dark matter and phantom field. Further $\mathscr{L}_{I}$ can be regarded as negligible because of small interaction between the phantom field and the dark matter. 
Let us then write the Einstein field  equation for the Einstein-Maxwell gravity in the presence of PFDM,   
\begin{align}
G_{\mu\nu}=R_{\mu\nu}-\frac{1}{2}g_{\mu\nu}R&=T_{\mu\nu}\ ,\label{eq:N3}\\
\nabla_{\mu}F^{\mu\nu}&=0\ ,\label{eq:N4}
\end{align}
where the energy-momentum tensor $T_{\mu\nu}$ is defined by 
\begin{align}
T_{\mu\nu}&=2F_{\mu\alpha}F_{\nu}^{~\alpha}-\frac{1}{2}g_{\mu\nu}F_{\alpha\beta}F^{\alpha\beta}\nonumber\\
&+2\nabla_\mu\Phi\nabla_\nu\Phi-g_{\mu\nu}\nabla_\alpha\Phi\nabla_\alpha\Phi+\nonumber\\&+T_{\mu\nu}^{\rm DM}+T_{\mu\nu}^{\rm I}\ ,
\end{align}
with $T_{\mu\nu}^{\rm DM}$ representing the energy-momentum tensor for the {PFDM}.

We consider the static, spherically symmetric metric ansatz for the black hole as
\begin{equation}
ds^2=-e^{A(r)}dt^2+e^{B(r)}dr^2+r^2\left(d\theta^2+\sin^{2}\theta d\phi^2\right)\, ,
\end{equation} 
and then the Einstein equations with Maxwell field 
can be written in the form \cite{Li-Yang12}
\begin{align}
&e^{-B}\left(\frac{1}{r^2}-\frac{B'}{r}\right) -\frac{1}{r^2}=\frac{Q^2}{r^4}+\frac{1}{2}e^{-B}\Phi'^2-V(\Phi)-\rho_{\rm DM}\ ,\label{eq:einstein1}
\\
&e^{-B}\left(\frac{1}{r^2}+\frac{A'}{r}\right) -\frac{1}{r^2}=\frac{Q^2}{r^4}-\frac{1}{2}e^{-B}\Phi'^2-V(\Phi)\ ,\label{eq:einstein2}
\\
&e^{-B}\left(A''+\frac{A'^2}{2}+\frac{A'-B'}{r}-\frac{A'B'}{2}\right)=-\frac{Q^2}{r^4}+\frac{1}{2}e^{-B}\Phi'^2\nonumber\\&-V(\Phi)\ .\label{eq:einstein3}
\end{align}

It is worth noting that the components $\theta\theta$ and $\phi\phi$ of the Einstein field equations are identical.
From the above Einstein equation, the black hole metric in Boyer-Lindquist coordinates $x^\alpha=(t,r,\theta,\phi)$ is then given by the following line element:
\begin{align}\label{eq:metric1}
ds^{2}=-F(r)dt^2+\frac{dr^{2}}{F(r)}+r^2\left(d\theta^{2}+\sin^{2}\theta d\phi^2\right),
\end{align}
where $F(r)$ is defined as
\begin{align}\label{eq:F}
F(r)=1-\frac{2M}{r}+\frac{Q_{e+m}^2}{r^2}
+\frac{\lambda}{r}  \ln\frac{r}{\vert\lambda \vert} \ ,
\end{align}
with $M$ being the black hole mass, $Q_{e+m}^2=Q^{2}_{e}+Q^{2}_{m}$ where $Q_{e}$ and $Q_{m}$ are the electric and magnetic charges and $\lambda$ is related to the scale parameter characterized by the {PFDM}. {Thus, the static and spherically symmetric solution black hole can be endowed with three parameters, i.e. mass $M$, the electric charge $Q_e$ and magnetic charge $Q_m$. 
Note that parameters, $M$, $Q_{e,m}$ and $\lambda$ are dimensionful quantities, and hence their dimensions are taken to be $L^{1}$ having set $G=c=1$. However, for further analysis we shall for
simplicity normalize these parameters as $Q_{e,m}/M$ and $\lambda/M$, respectively, in order to define them as dimensionless quantities having set $M=1$.} For $Q_{e}=Q_{m}=\lambda=0$, the black hole spacetime metric surrounded by the PFDM field, i.e., Eq.~(\ref{eq:metric1}), reduces to the Schwarzschild metric, while  for $Q_{m}=\lambda=0$ the spacetime metric then reduces to the Reissner-Nordstr\"{o}m metric. Similarly, in the case of vanishing $Q_{e}$ and $Q_{m}$, it reduces to the Schwarzschild black hole surrounded by {PFDM} field. {In the case of nonvanishing dark matter distribution, i.e. $\lambda\neq 0$, the energy-momentum tensor for an anisotropic perfect fluid distribution is written as 
\begin{align}
T^\mu_\nu={\rm diag}(-\rho,p_{r},p_{\theta},p_{\phi})\, .
\end{align}
Here density, radial and tangential pressures will respectively read as 
\begin{align}
\rho=-p_{r}= \frac{\lambda}{8\pi r^3}\,  \mbox{~~~and~~~} p_{\theta}=p_{\phi}=\frac{\lambda}{16\pi r^3}\, .\label{eq:rho}
\end{align}
We shall only consider the positive value for a dark matter profile $\lambda>0$ for further analysis, which refers to positive energy density that represents attractive behavior.}

The horizon of the black hole is determined as a solution of the following nonlinear equation,
\begin{align}\label{eq:hor}
1-\frac{2M}{r}+
{\frac{Q^2_{e+m}}{r^2}}=-\frac{\lambda}{r}\ln\frac{r}{|\lambda|}\, .    
\end{align}
In the case of vanishing $\lambda$ the horizon radius takes the simple form as
\begin{align}\label{eq:4}
r_{\pm}=M\pm\sqrt{M^{2}-
{Q^2_{e+m}}}.
\end{align}
The above equation corresponds to the outer and inner horizons. If the two outer and inner horizons coincide, i.e. $r_{+}=r_{-}$, it is then interpreted by the extremal charged dyonic black hole for which the following condition is satisfied 
\begin{align}\label{eq:5}
M^{2}= 
{Q^2_{e+m}},
\end{align}
while the black hole horizon no longer exists in the case of $M^{2}< 
{Q^2_{e+m}}$, exhibiting a naked singularity. 

\begin{figure}
    \centering
    \includegraphics[scale=0.5]{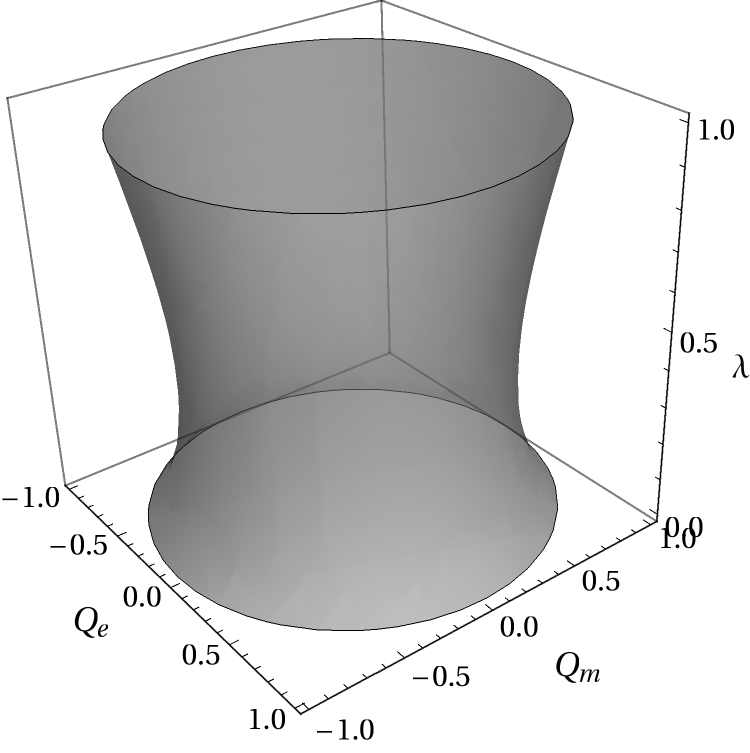}
    \caption{The limits on the parameters $Q_e, Q_m$, and $\lambda$ are presented. The shaded region corresponds to the permitted region for which a black hole exists, while the unfilled region corresponds to the naked singularity.}
    \label{fig:QeQmlambda_region_plt}
\end{figure}

\begin{figure}
    \centering
    \includegraphics[scale=0.52]{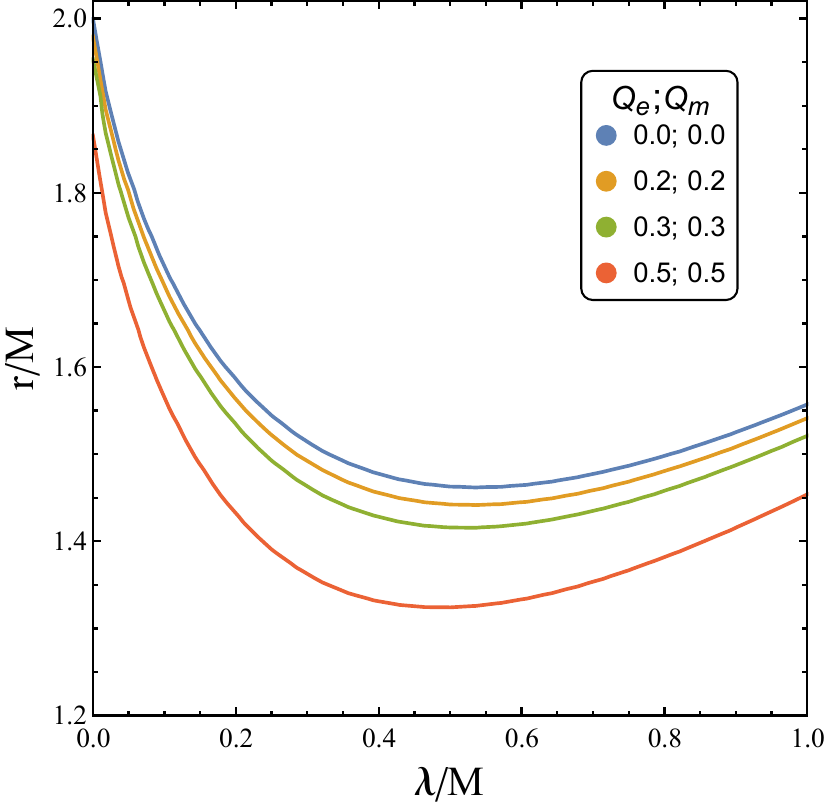}
    \caption{{The horizon radius $r_h$ is plotted as a function of perfect fluid dark matter parameter $\lambda$ for various values of $Q_e$ and $Q_m$.}}
    \label{fig:hor_dm}
\end{figure}

In the case of nonvanishing $\lambda$, it becomes complicated to solve the nonlinear \cref{eq:hor} analytically for the black hole horizon. It is, however, possible to reach its analytic form by imposing the condition, i.e., $\lambda \ll M$. In doing so, one can get the horizon radius in the following form 
\begin{eqnarray}
r_{\pm}&=& M\pm\Bigg[M^2-
{Q^2_{e+m}}\nonumber\\&&-\lambda \left(M+\sqrt{M^2-
{Q^2_{e+m}}}\right) \nonumber\\ &&\times  \ln \left(\frac{M+\sqrt{M^2-
{Q^2_{e+m}}} 
}{\lambda}\right)\Bigg]^{1/2}\, .
\end{eqnarray}
Hereafter, we only use the black hole's outer horizon $r_{+}$ popularly known as the event horizon, and use the notation $r_{h}$ instead of $r_{+}$, unless otherwise specified. 
In Fig.~\ref{fig:QeQmlambda_region_plt}, we show the limits on the parameters $Q_e$, $Q_m$, and $\lambda$. As can be seen from Fig.~\ref{fig:QeQmlambda_region_plt}, the shaded region corresponds to the permitted region for which a black hole exists, while the unfilled region corresponds to the naked singularity. {To understand a bit more clearly, in Fig.~\ref{fig:hor_dm}, we show the dependence of the horizon radius $r_h$ on the {PFDM} parameter {$\lambda$}. From Fig.~\ref{fig:hor_dm}, the black hole horizon radius decreases as parameters $Q_e$, $Q_m$, and $\lambda$ increase till some particular value of $\lambda$, and then it starts to grow. This clearly shows that the parameter {$\lambda$} has the physical effect that shifts black hole's outer horizon toward to small $r$. This is in agreement with the physical meaning of the parameter $\lambda$ as a repulsive gravitational charge, similarly to the black hole's electric and magnetic charges $Q_{e,m}$. However, it can be seen from Fig.~\ref{fig:hor_dm} that the radius of the horizon grows in the limit of large $\lambda$. This happens because the radial pressure $p_r$ (see, \cref{eq:rho}) has the repulsive nature, and thus it turns out that the dark matter field, $\rho$, can be repulsive in the limit of large $\lambda$. Hence, we further focus on the small $\lambda$ in order to manifest more realistic model for the dark matter distribution.} 

{In the following sections, the motion of magnetically and electrically charged particles, as well as that of neutral spinning particles will be studied. It's worth noting that, we use the Hamiltonian technique to investigate the motion of magnetically and electrically charged particles just because of simplicity. On the other hand to investigate the motion of spinning particle, there are two well-known approaches mainly: the Mathisson-Papapetrou (MP) approach \cite{Mathisson:1937zz,Papapetrou:1951pa} and the Lagrangian approach \cite{Hanson:1974qy,Hojmanphdthesis:1975,Hojman:1976kn,Hojman:2012me}. For the spinning particle, we use the Lagrangian technique rather than the MP approach in this article. The rationale for adopting the Lagrangian approach over the MP approach is explained in \cref{Sec:Appendix_A}
}

\section{Charged particle dynamics\label{Sec:motion}}	

{First,} we focus on charged particle motion around a charged black hole in the presence of perfect fluid dark matter. We assume that the test particle is endowed with the rest mass $m$, electric charge $q$, and magnetic charge $q_m$. In general, the Hamilton-Jacobi equation of the system is then expressed as \cite{Misner73}
\begin{align}
H&=g^{\alpha\beta}\left(\frac{\partial S}{\partial x^\alpha}-qA_\alpha+iq_mA_\alpha^{\star}\right)\nonumber\\&\times \left(\frac{\partial S}{\partial x^\beta}-qA_\beta+iq_mA_\beta^{\star}\right)\, ,\label{eq:HJ}  
\end{align}
with the action $S$ with the spacetime coordinates, the components of the vector and the dual vector potentials $A_\alpha$ and $A_\alpha^{\star}$, and electric and magnetic charges $q$ and $q_m$. The components of the associated vector potentials $A_\alpha$ and $A_\alpha^{\star}$ of the electromagnetic field take the following form 
\begin{align}
A_{\alpha}&=\left(-\frac{Q_{e}}{r},0,0,Q_{m}\cos\theta\right)\, ,\nonumber\\
A_{\alpha}^{\star}&=\left(-\frac{iQ_{m}}{r},0,0,iQ_e\cos\theta\right)\, .\label{eq:vec_potential_A}
\end{align}
As one can see, the spacetime \cref{eq:metric1} and the components of the vector potential \cref{eq:vec_potential_A} are independent of coordinates ($t, \phi$) which lead to two conserved quantities, namely, energy $E$, and angular momentum $L$ of the charged particle measured at infinity. It is known that the Hamiltonian is regarded as a constant $H=k/2$ with relation to $k=-m^2$, where $m$ is the 
mass of a particle having electric and magnetic charges. Following the Hamilton-Jacobi equation for charged particle motion, we write the action $S$ as follows 
\begin{align}\label{eq:Sol}
S=-\frac{1}{2}k\lambda -Et+L\phi+S_r+S_\theta\ ,
\end{align}
where $S_r$ and $S_\theta$, respectively, refer to the radial and angular functions of only $r$ and $\theta$. 
Using \cref{eq:Sol} one can easily obtain the Hamilton-Jacobi equation in the following form 
\begin{align}\nonumber
&\left(\frac{\partial S_\theta}{\partial\theta}\right)^2+\left(\frac{L-qQ_m\cos\theta}{\sin\theta}\right)^2\\&=\frac{r^2}{F}\left(E+\frac{qQ_e}{r}-\frac{q_mQ_m}{r}\right)^2-r^2F\left(\frac{\partial S_r}{\partial r}\right)^2+kr^2\ .\label{eq:HJ1}    
\end{align}
Here one can see that \cref{eq:HJ1} is fully separable into radial and angular parts. Hereafter, performing simple algebraic manipulations, one can show that
\begin{align}
&S_r=\int\frac{dr}{F}\sqrt{\left(E+\frac{qQ_e}{r}-\frac{q_mQ_m}{r}\right)^2-F\left(-k+\frac{K}{r^2}\right)}\ , \label{HJ1} \\ &S_\theta=\int d\theta\sqrt{K-\left(\frac{L-qQ_m\cos\theta}{\sin\theta}\right)^2}\ , \label{eq:HJ2}
\end{align}
where $K$ is the Carter constant of motion.

From the above equation, $E$, $L$, $k$, and $K$ are independent constants of motion. The fourth one is related to the latitudinal motion and caused by the separability of the action. If we focus on the
equatorial motion (i.e., $\theta=\pi/2$) we can further eliminate the fourth constant of motion \cite{Misner73}.  We shall for convenience introduce the following notations:
\begin{eqnarray}
{\cal E}&=&\frac{E}{m}\, ,~ {\cal L}=\frac{L}{mM}\, \mbox{~~and~~} 
~\frac{k}{m^2}=-1\ .
\end{eqnarray}
%

%
%
%

Following \crefrange{eq:HJ1}{eq:HJ2}, we obtain the radial equation of motion for charged particles in the following form
\begin{align}\label{Eq:rdot}
\dot{r}^2=\Big(\mathcal{E}-
{V_{eff(+)}(r)}\Big)\Big(\mathcal{E}-
{V_{eff(-)}(r)}\Big)\, ,
\end{align}
where $
{V_{eff(\pm)}(r)}$ {are positive and negative solution of the effective potential $V_{eff}(r)$ describing} the radial function of the radial motion and is given by   
\begin{align} \label{eq:Veff1}
{V_{eff(\pm)}(r)}&= \frac{ g_m}{r}-\frac{ \sigma_e}{r} \pm
\left(1+\frac{\mathcal{L}^2}{r^{2}}\right)^{1/2}\nonumber\\ &\times \left(1-\frac{2M}{r}+
{\frac{ Q^2_{e+m}}{r^2}}+\frac{\lambda}{r}\ln\frac{r}{\vert\lambda \vert}\right)^{1/2}\, , 
\end{align}
where the charge coupling parameters are defined as
\begin{align}
\sigma_e=\frac{qQ_e}{mM}\qquad
\mbox{and}\qquad g_m=\frac{q_mQ_m}{mM}\, , 
\end{align}
and in further calculations the radial coordinate and dark matter parameter are normalized as $r\to r/M$ and $\lambda \to \lambda /M$, respectively.
As seen from \cref{Eq:rdot},  we have either $\mathcal{E}>
{V_{eff(+)}(r)}$ or $\mathcal{E}<
{V_{eff(-)}(r)}$ since $\dot{r}^2\geq 0$ always. However, we select $
{V_{eff(+)}(r)}$ as an effective potential,  i.e. $V_{ eff}(r)=
{V_{eff(+)}(r)}$.
{From \cref{eq:Veff1}, if we remove all parameters except black hole mass $M$, we can simply recover the effective potential as in the Schwarzschild spacetime}.  
Also, we note that we have written the radial motion for a charged test particle in the equatorial plane (i.e., $\theta=\pi/2$) in the form given by \cref{eq:Veff1}.
\begin{figure*}
\begin{tabular}{c c }
  \includegraphics[scale=0.6]{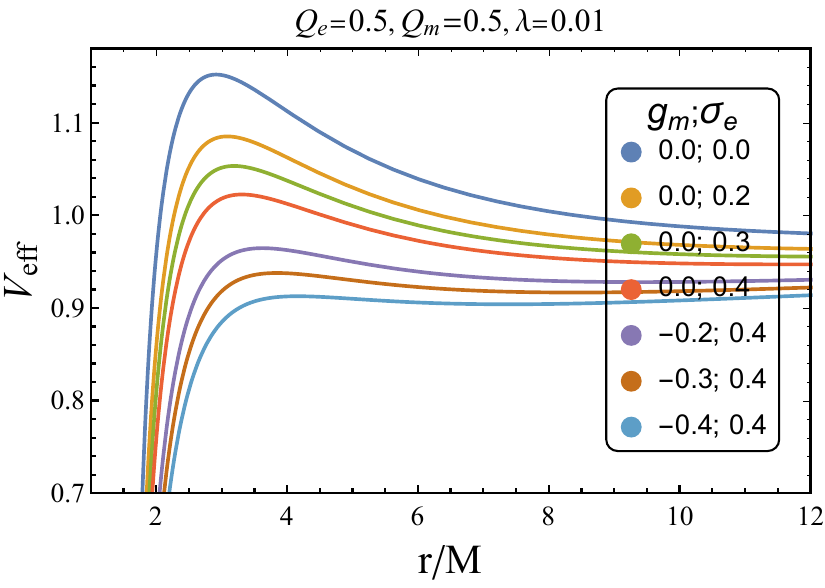}\hspace{0cm}
  &  \includegraphics[scale=0.6]{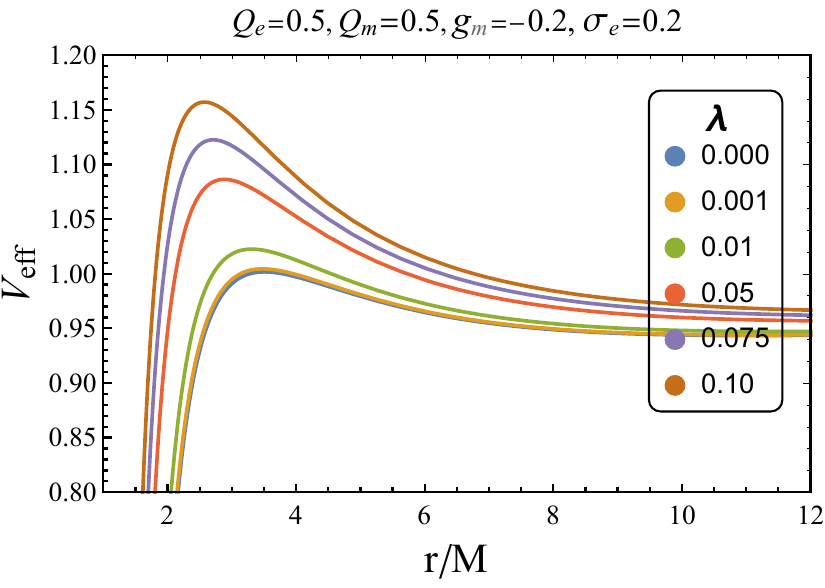}
    \end{tabular}
\begin{tabular}{c }
      \includegraphics[scale=0.6]{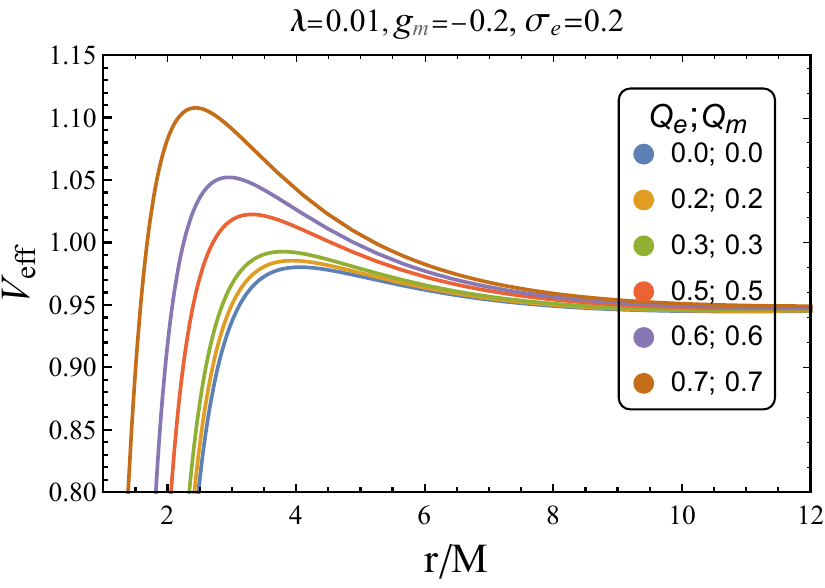}
\end{tabular}
	\caption{\label{fig:Veff_charged} 
Radial profile of the effective potential $V_{eff}$ as a function of the spatial distance $r/M$ for massive charged particles orbiting an electrically and magnetically charged black hole immersed in PFDM field. Top row, left panel: $V_{eff}$ is plotted for different values of the charge coupling parameters $g_m$ and $\sigma_e$. Top row, right panel: $V_{eff}$ is plotted for different values of dark matter parameter $\lambda$. The panel in the bottom row shows $V_{eff}$ versus $r/M$ for different values of black hole electric $Q_e$ and magnetic charge $Q_m$ while keeping fixed the dark matter parameter $\lambda$ and the charge coupling parameters $g_m$ and $\sigma_e$.}
\end{figure*}
%
\begin{figure*}
\centering
\includegraphics[width=1.0\textwidth]{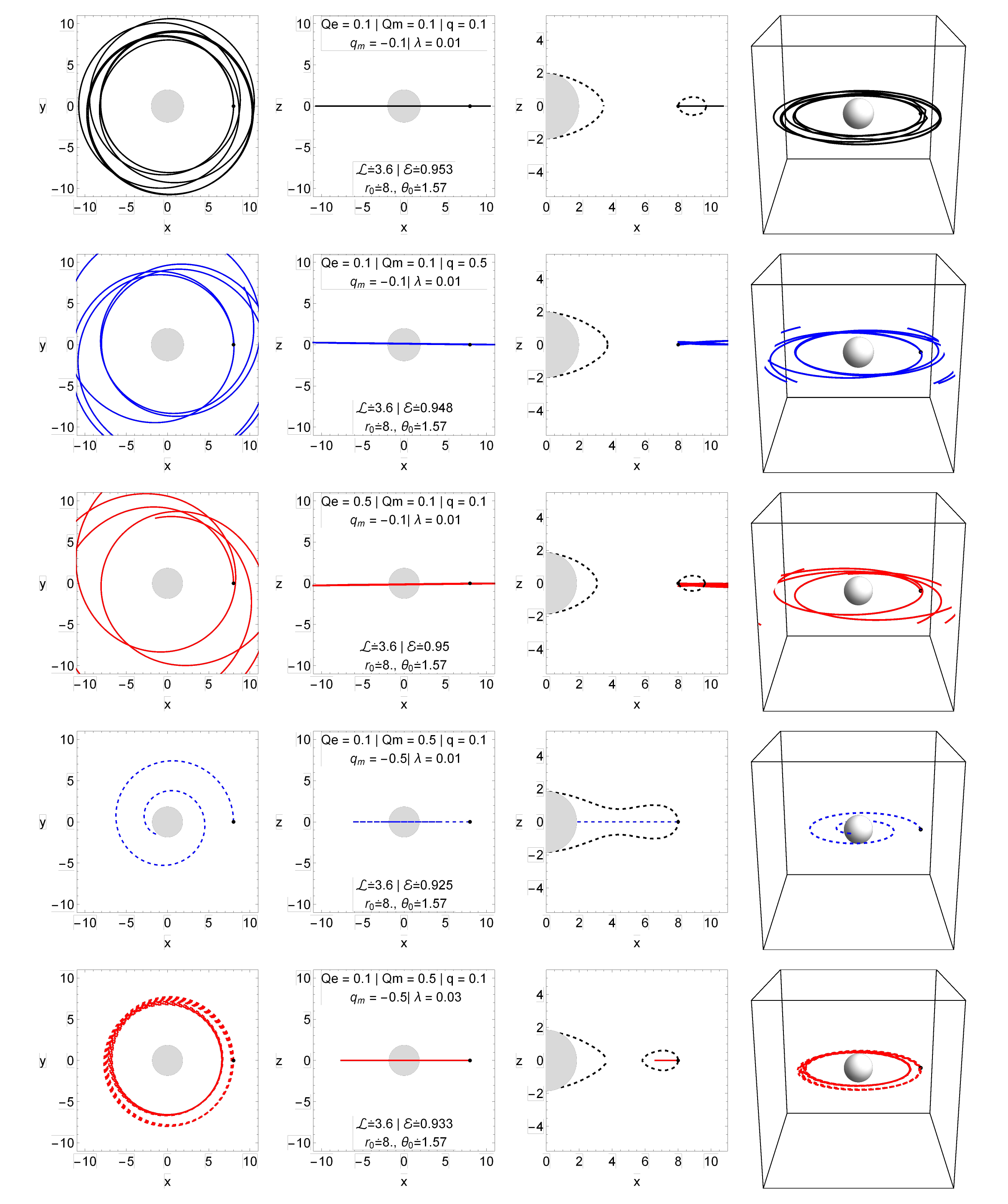}%
\vspace{-0cm}

\caption{\label{fig:traj} 2D and 3D trajectories of the charged particles observed from the polar view (i.e., $z = 0$; see the first column) and from the equatorial view (i.e., $y = 0$; see the second column), as well as the boundaries of the motion of charged particles observed from the equatorial view (i.e., $y = 0$; see the third column) around an electrically and magnetically charged dyonic black hole surrounded by PFDM field for various possible cases. 
} 
\end{figure*}
\begin{figure*}
\begin{tabular}{c c}
  \includegraphics[scale=0.6]{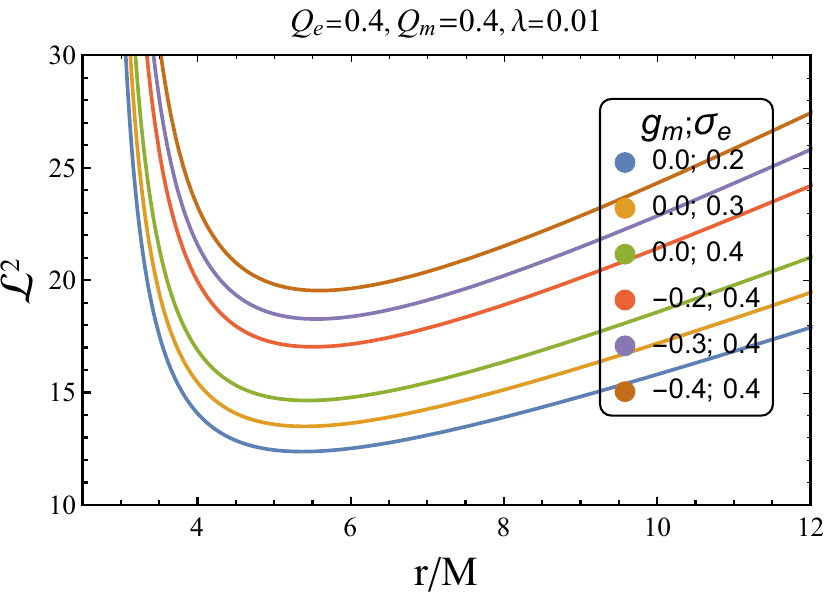}\hspace{0cm}
  &  \includegraphics[scale=0.6]{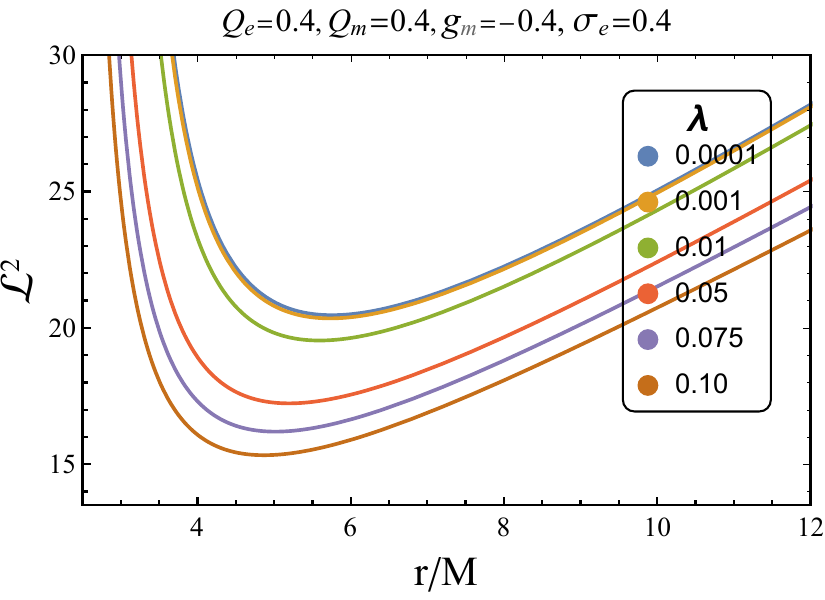}
\end{tabular}
	\caption{\label{fig:ang} 
Radial profile of the specific angular momentum $\mathcal{L}$ as a function of the spatial distance $r/M$ for massive charged particles orbiting a charged dyonic black hole immersed in PFDM field. {Left panel:} $\mathcal{L}^2$ is plotted for different values of charge coupling parameters $g_m$ and $\sigma_e$ while keeping the {PFDM} parameter $\lambda$ and $Q_e$ and $Q_m$ fixed. {Right panel:} $\mathcal{L}^2$ is plotted for different values of {PFDM} parameter $\lambda$ while keeping the rest of the parameters fixed as shown in the plot. }
\end{figure*}
%


{
In an astrophysical scenario, it is believed that the environment surrounding black holes cannot be regarded as vacuum as a consequence of the existence of matter and fields. Therefore, it would become increasingly important to take into account the repulsive and attractive effects due to the matter distribution in nearby environment of the black hole. For that we analyze the radial profile of the effective potential $V_{eff}(r)$ that reflects the 
{trajectories} of particles for various combinations of charge coupling parameters ($\sigma_{e}$ and $g_{m}$) and black hole parameters involving black hole charges ({$Q_{e,M}$})  and dark matter parameter {$\lambda$} that describe the background geometry.}
As can be seen from Fig.~\ref{fig:Veff_charged}, the left panel in the top row reflects the role of 
{$\sigma_{e}$ and $g_{m}$} for the radial profile of the effective potential, while keeping black hole charges $Q_e$, $Q_m$ fixed  and dark matter parameter $\lambda$, {whereas}  the right panel reflects the role of parameter $\lambda$ {for the case when parameters $Q_{e}, Q_{m}, \sigma_{e}$, and $g_{m}$ are fixed}. The panel in the bottom row of Fig.~\ref{fig:Veff_charged} shows the impact of the {parameters $Q_{e}$ and $Q_{m}$} on the radial profile of $V_{eff}(r)$ 
{for the case when parameters $\lambda, \sigma_{e}$ and $g_{m}$ are fixed}.  From \cref{fig:Veff_charged}, the height and strength of $V_{eff}(r)$ increase, and its shape shifts toward {the event horizon $r_{h}$}  as a consequence of the impact of {black hole parameters $Q_{e}, Q_{m}$, and $\lambda$}. 
{Further, examination of \cref{fig:Veff_charged} shows that the inclusion of parameters $\sigma_{e}$ and $g{m}$ reduces the maximum of $V_{eff}(r)$ and pushes it away from the horizon $r_{h}$.
} {Also}, one can infer that the combined effect of parameters {$Q_{e,m}$ and $\lambda$} on the profile of the {$V_{eff} (r)$} is balanced by the parameters {$\sigma_{e}$ and $g_{m}$}. As was stated earlier, we restrict the motion to the equatorial plane, i.e., $\pi=\theta/2$. 
{
We then further find a particle trajectory that helps us to understand qualitatively how charged test particle behaves around the black hole under the combined effect of electromagnetic and gravitational forces. In \cref{fig:traj}, we show particle trajectory restricted to move on the equatorial plane of the charged dyonic black hole immersed in PFDM. 
As seen in \cref{fig:traj}, there occur for various orbits, i.e. bound, captured and the escaping orbits, depending upon the parameter choices. It is obvious from the particle trajectory that orbits are captured by black hole when the magnetic charge $q_{m}$ {of the particle} increases, while they turn the escaping orbits for larger electric charge $q_{e}$ {of the particle}. However, it turns out that these orbits become bounded as a consequence of the increase in value of PFDM {parameter $\lambda$} in the case of fixed particle's energy and angular momentum as shown in {sub} plots {of \cref{fig:traj}}.}

Now we turn to the study of the circular orbits of electrically and magnetically charged test particles around the charged dyonic black hole in the presence of PFDM. {As we know from the theory of geodesic particles that a test particle needs to satisfy following conditions simultaneously in order to move in circular orbits} 
{
\begin{itemize}
   \item the radial velocity should vanish:
    \begin{align}
        \frac{dr}{d\tau}=0 \;\implies\; \mathcal{E}=V_{eff}(r).\label{eq:dotr11} 
    \end{align}
   \item the radial acceleration should also vanish:
   \begin{align}
       \frac{d^{2}r}{d\tau^{2}}=0 \;\implies\;\frac{d V_{eff}(r)}{dr}=0 .\label{eq:ddotr1}
   \end{align}
\end{itemize}
These two conditions determine the circular orbits both types of circular orbits i.e. stable and unstable.}

%
The above equations lead us to determine the specific angular momentum $\mathcal{L}$ and energy $\mathcal{E}$ for electrically and magnetically charged particles on the circular orbits
%
\begin{align}
\mathcal{L}^2&= \frac{2 F(r) \left[r^3 F'(r)+(g_m-\sigma_e )^2\right]-r^4 F'(r)^2}{\Big(r F'(r)-2 F(r)\Big)^2}\nonumber\\&+ \frac{2F(r)  \sqrt{4 r^2 F(r)-2 r^3 F'(r)+(g_m-\sigma_e )^2}}{(g_m-\sigma_e )^{-1}\Big(r F'(r)-2 F(r)\Big)^2}
 \, , \label{eq:L_charge}\\
 \mathcal{E}&= \frac{ g_m}{r}-\frac{ \sigma_e}{r} +
\left(1+\frac{\mathcal{L}^2}{r^{2}}\right)^{1/2}\nonumber\\ &\times \left(1-\frac{2}{r}+
{\frac{Q^{2}_{e+m}}{r^{2}}}+\frac{\lambda}{r}\ln\frac{r}{\vert\lambda \vert}\right)^{1/2}\, , \label{eq:E_charge}
\end{align}
%
where $^{\prime}$ refers to a derivative with respect to $r$. It is worth noting that in the case $g_m=\sigma_e$, we obtain $\mathcal{L}$ for the Reissner-Nordstr\"{o}m black hole case. We now analyze the radial profiles of the specific angular momentum for charged particles so as to understand the behavior of the circular orbits around the black hole. In \cref{fig:ang}, we show the radial profile of the specific angular momentum $\mathcal{L}$ for charged particles at the  circular orbits  around 
{ a charged dyonic} black hole immersed in PFDM. As shown in Fig.~\ref{fig:ang}, the left panel reflects the impact of the charge coupling parameters $g_m$ and $q_e$ on the radial profile of $\mathcal{L}$ for fixed values of black hole charges $Q_e$ and $Q_m$ and {PFDM} parameter $\lambda$, while the right panel reflects the impact of the dark matter while keeping the black hole charge and charge coupling parameters fixed. 
From \cref{fig:ang}, {one can easily see that circular orbits of electrically and magnetically charged particles move at larger radii as we increase the charge coupling parameters $g_m$ and $\sigma_e$ (see left panel)} 
In contrast, circular orbits move {toward} smaller radii and become close to the black hole {horizon $r_{h}$} with increasing {PFDM} parameter $\lambda$. Additionally, what we observe from the above analysis is that both black hole charges $Q_{e,m}$ and {PFDM parameter $\lambda$} bring the radii of the circular orbits closer to the {$r_{h}$} of the black hole, thus behaving as a repulsive nature as a result 

On the other hand, another limit of the existence of circular orbits is described by the radius for unstable circular photon orbits {$r_{ph}$}, which is determined by the divergence of either the specific angular momentum {$\mathcal{L}$} or energy {$\mathcal{E}$}. From \crefrange{eq:L_charge}{eq:E_charge}, one can easily obtain the limit on the existence of circular orbits that always exist at $r>r_{ph}$, %
\begin{eqnarray}\label{Eq:ph} 
r F'(r)-2 F(r)=0\, ,
\end{eqnarray} 
 which gives the radii of the photon orbits. We shall, for simplicity, consider $\lambda \ll 1$ to determine the approximate expression for the photon orbit $r_{ph}$ as   
{\begin{align}\label{Eq:ph1}
r_\text{ph}&\approx  \frac{1}{2} \Big(3M+\Big[9M^2-
{8Q_{e+m}^2}\nonumber\\&+  \lambda \left(3M+\sqrt{9M^2-
{8Q_{e+m}^2}}\right)\nonumber\\&- 3 \lambda \left(3M+\sqrt{9M^2-
{8Q_{e+m}^2}}\right) \nonumber\\&\times \ln \left(\frac{3M+\sqrt{9M^2-
{8Q_{e+m}^2}}}{2 \lambda}\right)\Big]^{1/2}\Big) \nonumber\\&+ O(\lambda^2) \, . 
\end{align}}%
From the above expression one can easily see that it explicitly {reduces to the Schwarzschild case,}  i.e., $r_{ph}=3$ in the limit $\lambda, Q_{e,m}\to 0$. However, it is certain {from \cref{Eq:ph1}} that $r_{ph}$ decreases as we increase both the parameters $\lambda$ and {$Q_{e,m}$}. 

%
{Next, in order to determine the ISCO we need to find limiting value of orbital angular momentum $\mathcal{L}$ for which $V_{eff}(r)$ still has an extremum. The ISCO lies at the position where the maximum and minimum of effective potential $V_{eff(r)}$ merge. Hence, one more condition is needed beside conditions \crefrange{eq:dotr11}{eq:ddotr1}, which reads
\begin{itemize}
    \item  the second derivative of $V_{eff(r)}$ with respect to radial coordinate $r$ should vanish:
    \begin{align}       \frac{d^{2}V_{eff(r)}}{dr^{2}}=0.\label{eq:ddVeff}
    \end{align}
\end{itemize}}
For being somewhat more quantitative, in \cref{tab:table1} we show the numerical values of 
{ISCO parameters $\mathcal{L}_{ISCO}, \mathcal{E}_{ISCO}$ and $r_{ISCO}$ for various combinations of $Q_{e,m}, \sigma_{e}, g_{m}$ and $\lambda$}. As shown in \cref{tab:table1} 
{both $Q_{e,m}$ and $\lambda$} have similar effect that decreases the {$r_{ISCO}$}. However, the {$r_{ISCO}$} increases as a consequence of the increasing the charge coupling parameters $g_m$ and $\sigma_e$. Furthermore, one can see that the combined effect of {both $Q_{e,m}$ and $\lambda$} respectively decreases 
{$\mathcal{L}_{ISCO}$ and increases $\mathcal{E}_{ISCO}$}, while the opposite is the case as we increase the charge coupling parameters, as seen in \cref{tab:table1}.     
\begin{table*}
\begin{center}
\caption{{Numerical values of the 
{ISCO parameters $\mathcal{L}_{ISCO}, \mathcal{E}_{ISCO}, r_{ISCO}$, $v_{ISCO}$ and $\Omega_{ISCO}$} of the test particles orbiting on the ISCO around the {charged dyonic} black hole surrounded by PFDM field are tabulated for various possible cases.}}\label{tab:table1}
\resizebox{\linewidth}{!}
{
\begin{tabular}{l l |l l l l l| l l l l l| l l l l l}
 \hline \hline
 & & &\multicolumn{2}{c|}{$Q_{e}/Q_{m}=0.0/0.0$} & $\sigma_e/g_{m}=0.0/0.0$ & &\multicolumn{2}{c|}{$Q_{e}/Q_{m}=0.4/0.0$} & $\sigma_e/g_{m}=0.0/0.0$ & &  &\multicolumn{2}{c|}{$Q_{e}/Q_{m}=0.4/0.0$}& $\sigma_e/g_{m}=0.2/0.0$\\
\cline{3-7}\cline{8-12}\cline{13-17}
& $\lambda$    & $\mathcal{L}_{ISCO}$ & $\mathcal{E}_{ISCO}$ & $r_{ISCO}$ & $v_{ISCO}$ & $\Omega_{ISCO}$ & $\mathcal{L}_{ISCO}$ & $\mathcal{E}_{ISCO}$ & $r_{ISCO}$ & $v_{ISCO}$ & $\Omega_{ISCO}$ & $\mathcal{L}_{ISCO}$ & $\mathcal{E}_{ISCO}$ & $r_{ISCO}$ & $v_{ISCO}$ & $\Omega_{ISCO}$ \\
\hline
& 0.00001   & 3.46388 & 0.94280  & 5.99964 & 0.49999 & 0.06804 & 3.38450 & 0.94056  & 5.75241 & 0.50710 & 0.07146 & 3.70498 & 0.92948 & 5.78308 & 0.50510 & 0.07090\\
& 0.0001   & 3.46231 & 0.94281  & 5.99709 & 0.49998 & 0.06807 & 3.38289 & 0.94057  & 5.74975 & 0.50709 & 0.07149 & 3.70337 & 0.92949 & 5.78043 & 0.50509 & 0.07093\\
& 0.001     & 3.45019 & 0.94288 & 5.97791 & 0.49987 & 0.06828 & 3.37054 & 0.94063 & 5.72974 & 0.50703 & 0.07174 &3.69097 & 0.92950 & 5.76035 &0.50503 & 0.07117\\
& 0.01       & 3.36533 & 0.94361 & 5.84896 & 0.49871 & 0.06973 & 3.28414 & 0.94127 & 5.59512 & 0.50620 & 0.07344 & 3.60415 & 0.92986 & 5.62457 &0.50423 & 0.07287\\
& 0.1         & 2.89896 & 0.95197 & 5.22562 & 0.48511 & 0.07727 & 2.81143 & 0.94926 & 4.94006 & 0.49461 & 0.08260 &3.12739 & 0.93595 & 4.94999 & 0.49386 & 0.08235\\
& 0.2         & 2.60524 & 0.96260 & 4.93536 & 0.46682 & 0.08052 & 2.51714 & 0.95991 & 4.62883 & 0.47773 & 0.08703 &2.82894 & 0.94510 & 4.60552 & 0.47959 & 0.08771\\
\hline
& &&\multicolumn{2}{c|}{$Q_{e}/Q_{m}=0.4/0.4$}&  $\sigma_e/g_{m}=0.2/-0.2$ && \multicolumn{2}{c|}{$Q_{e}/Q_{m}=0.4/0.4$} & $\sigma_e/g_{m}=0.4/-0.4$ & &  &\multicolumn{2}{c|}{$Q_{e}/Q_{m}=0.4/0.4$}  & $\sigma_e/g_{m}=0.6/-0.6$\\
\cline{3-7}\cline{8-12}\cline{13-17}
& $\lambda$    & $\mathcal{L}_{ISCO}$ & $\mathcal{E}_{ISCO}$ & $r_{ISCO}$ & $v_{ISCO}$ & $\Omega_{ISCO}$ & $\mathcal{L}_{ISCO}$ & $\mathcal{E}_{ISCO}$ & $r_{ISCO}$ & $v_{ISCO}$ & $\Omega_{ISCO}$ & $\mathcal{L}_{ISCO}$ & $\mathcal{E}_{ISCO}$ & $r_{ISCO}$ & $v_{ISCO}$ & $\Omega_{ISCO}$ \\
\hline
& 0.00001   & 3.93127 & 0.91596 & 5.59874 & 0.50781 & 0.07329 & 4.52552 & 0.89618 & 5.75601 &0.49774 & 0.07036 & 5.09500 & 0.87829 & 5.93189 &0.48716 & 0.06732\\
& 0.0001   & 3.92961 & 0.91595 & 5.59599 & 0.50780 & 0.07333 & 4.52384 & 0.89616 & 5.75328 &0.49773 & 0.07040 & 5.09329 & 0.87825 & 5.92918 &0.48715 & 0.06735\\
& 0.001     & 3.91685 & 0.91591 & 5.57510 & 0.50776 & 0.07360 & 4.51089 & 0.89605 & 5.73257 & 0.49764 & 0.07065 & 5.08012 & 0.87811 & 5.90859 & 0.48702 & 0.06757\\
& 0.01       & 3.82753 & 0.91584 & 5.43343 & 0.50720 & 0.07551 & 4.41997 & 0.89552 & 5.59141 & 0.49680 & 0.07239 & 4.98741 & 0.87719 & 5.76778 & 0.48591 & 0.06915\\
& 0.1         & 3.33649 & 0.91936 & 4.71908 & 0.49886 & 0.08667 & 3.91536 & 0.89583 & 4.86898 & 0.48756 & 0.08272 & 4.46823 & 0.87485 & 5.03933 &0.47559 & 0.07859\\
& 0.2         & 3.02976 & 0.92671 & 4.34079 & 0.48696 & 0.09372 & 3.59461 & 0.90046 & 4.47181 & 0.47621 & 0.08958 &4.13293 & 0.87719 & 4.62913 & 0.46415 & 0.08499\\
 \hline \hline
\end{tabular}
}
\end{center}
\end{table*}

Let us then consider the orbital and angular velocity of an electrically and magnetically charged particle moving around {the charged dyonic} black hole surrounded by PFDM, measured by a local observer~\cite{Shapiro83,Misner73}. For that, we first write the {coordinate} velocity components for the particles:
\begin{align}\label{eq:vr}
&v_{\hat r}=\sqrt{-\frac{g_{rr}}{g_{tt}}}\frac{dr}{dt}=\sqrt{1-F\frac{r^2+{\cal K}}{(r{\cal E}+\sigma_e-g_m)^2}}\ , \\
\label{eq:vt}
&v_{\hat\theta}=\sqrt{-\frac{g_{\theta\theta}}{g_{tt}}}\frac{d\theta}{dt}=\frac{\sqrt{F}}{r{\cal E}+\sigma_e-g_m}\sqrt{{\cal K}-\left(\frac{{\cal L}-\sigma_m\cos\theta}{\sin\theta}\right)^2}\ ,\\
&v_{\hat\phi}=\sqrt{-\frac{g_{\phi\phi}}{g_{tt}}}\frac{d\phi}{dt}=\frac{\sqrt{F}}{r{\cal E}+\sigma_e-g_m}\frac{{\cal L}-\sigma_m \cos\theta}{\sin\theta}\, ,\label{eq:vf}
\end{align}
where the Carter constant parameter $\mathcal{K}$ becomes irrelevant when we restrict motion to the equatorial plane, i.e., $\theta=\pi/2$.  
For the particle energy, its classical form can be written as follows:  
\begin{align}
{\cal E}=\frac{\sqrt{F}}{\sqrt{1-v^2}}-\frac{\sigma_e}{r}+\frac{g_m}{r}\, ,     
\end{align}
where $v=\left(v_{\hat r}^2+v_{\hat\theta}^2+v_{\hat\phi}^2\right)^{1/2}$ {and we have defined $v=\upsilon/c$.} 
As can be easily seen from Eqs.~(\ref{eq:vr}-\ref{eq:vf}) $v_{\hat\theta}=v_{\hat\phi}=0$ always, while the radial velocity is given by $v_{\hat r}=1$ very near the black hole horizon, i.e. $F(r)=0$. Similarly, one can determine the particle linear velocity at the ISCO radius (see, for example~\cite{Pugliese11}). The radial and latitudinal velocities vanish, i.e. $v_r=v_{\theta}=0$ at the ISCO, yet the particle orbital velocity takes the form 
\begin{align}\label{eq:v}
v_\phi=\sqrt{\frac{\partial_r g_{tt}}{\partial_r g_{\phi\phi}}\,\frac{g_{\phi\phi}}{g_{tt}}}\ ,    
\end{align}
where $\Omega=\sqrt{-\partial_r g_{tt}/\partial_r g_{\phi\phi}}$ corresponds to the orbital angular velocity of test particle, i.e., the so-called Keplerian angular frequency.
Equation~(\ref{eq:v}) then yields 
{\begin{align}\label{eq:v_r}
v=\sqrt{\frac{1}{2}\Big[\frac{r (\lambda+r)-Q_{e}^2-Q_{m}^2}{ r(r-2M)+Q_{e}^2+Q_{m}^2+\lambda\,r \ln \frac{r}{\vert\lambda\vert}}-1\Big]}\, .
\end{align}}
\begin{figure*}
\begin{tabular}{c c }
  \includegraphics[scale=0.6]{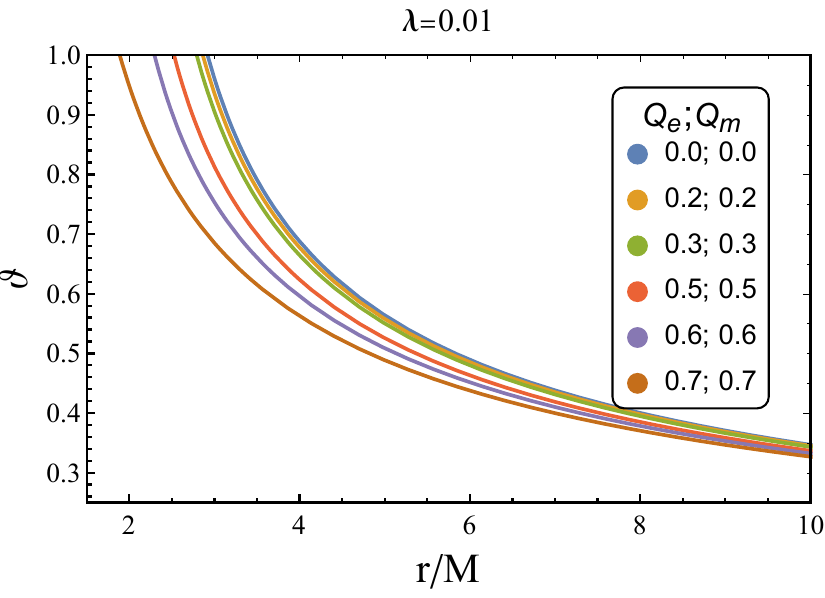}\hspace{0cm}
  &  \includegraphics[scale=0.6]{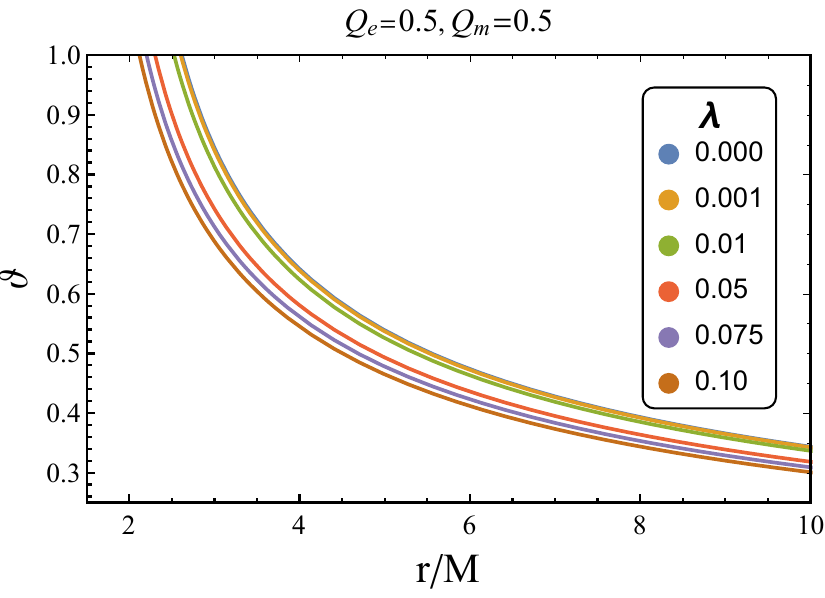}
\end{tabular}
	\caption{\label{fig:vel} 
Radial profile of the orbital velocity $v$ as a function of the spatial distance $r/M$ for test particles orbiting a black hole surrounded by PFDM field. {Left panel:} $v$ is plotted for different 
{combinations of black hole parameters $Q_e$ and $Q_m$} while keeping the {PFDM} parameter $\lambda$ {fixed}. {Right panel:} $v$ is plotted for different values of {PFDM} parameter $\lambda$ while keeping the black hole {charge parameters} $Q_e$ and $Q_m$ fixed. }
\end{figure*}
{It is worth noting here that the $v_{ISCO}$ for the Schwarzschild black hole (i.e. $v_{ISCO}=1/2$) may be readily recovered from the preceding equation when the black hole parameters $Q_{e,m}$  and $\lambda$ both disappear simultaneously.
} 

We now analyze the radial profile of orbital velocity $v$ for a test particle {at ISCO} in order to have detailed information about the background geometry of the {charged dyonic} black hole surrounded by PFDM as compared to the one for the Schwarzschild {black hole} case. In \cref{fig:vel}, we show the radial profile of the orbital velocity $v$ for test particles orbiting at the ISCO radius. From \cref{fig:vel}, one can easily see that the orbital velocity of test particle around {a charged dyonic} black hole immersed in PFDM decreases, and its shape shifts down to smaller $v$ as a consequence of an increase in the value of both black hole charges, {i.e., $Q_{e}$ and $Q_m$} and {PFDM} parameter $\lambda$. However, this may not be the case {in general}. Thus, we need to explore it numerically to understand the behavior of the orbital velocity better. The behavior of the results on the orbital velocity demonstrated in Fig.~\ref{fig:vel}, is explained in detail in Table \ref{tab:table1}. From \cref{tab:table1}, one can infer that the orbital velocity {$v$} decreases when increasing the {PFDM} parameter {$\lambda$},  while it increases due to the effect of black hole charges {$Q_{e}$ and $Q_{m}$}. 

\section{Spinning Particle Dynamics}
\label{sec:spinning_particles}
In this section, we focus on the {study of} motion of a spinning particle {moving} around {a charged dyonic} black hole in the presence of {PFDM}. Corinaldesi and Papapetrou \cite{Corinaldesi:1951pb} first studied the motion of a spinning particle moving in the vicinity of a Schwarzschild black hole. Later, Hojman \cite{Hojmanphdthesis:1975} used the Lagrangian approach to find the equation of motion for the spinning particles. {In this section, we use the Lagrangian approach to study the motion of a spinning particle (for a thorough explanation of why we chose the Lagrangian approach over the Mathisson \cite{Mathisson:1937zz} and Papapetrou \cite{Papapetrou:1951pa} approach, (see \cref{Sec:Appendix_A}).} 

We start by describing the Killing vectors for the metric of a charged dyonic black hole immersed in PFDM
\begin{align}
    \zeta^{0}_{\mu} &= (-F(r),0,0,0),\nonumber\\
    \zeta^{1}_{\mu} &= (0,0,0,r^{2}\sin^{2}{\theta}),\nonumber\\
    \zeta^{2}_{\mu} &= (0,0,r^{2}\sin{\phi},r^{2}\cos{\theta}\cos{\phi}\sin{\theta}),\nonumber\\
    \zeta^{3}_{\mu} &= (0,0,r^{2}\cos{\phi},-r^{2}\cos{\theta}\sin{\phi}\sin{\theta}).\label{eq:Killing_vectors}
\end{align}
As the metric (\ref{eq:metric1}) is independent of time, we can find the four constants of motion via a simple calculation using the above Killing vectors. For simplicity we are focusing only on the spinning particle motion in the equatorial (i.e., $\theta=\pi/2$) plane. First, we find the constant of motion (i.e., conserved energy $E$ and total angular momentum $J$  perpendicular to the $\theta=\pi/2$ plane) for the spinning particle using the Killing vectors and \cref{eq:Killing}
\begin{align}
    E &= F(r) P^{t}-\frac{1}{2}S^{tr}F(r)',\label{eq:conservationE}\\
    J &= r^{2}P^{\phi}+r S^{r\phi},\label{eq:conservationJ}
\end{align}
where the symbol ($'$) represents the derivative with respect to radial coordinate $``r"$.
Using the conservation \cref{eq:cons_P,eq:cons_S}, we find another two constants of motion (i.e., mass and spin): 
\begin{align}
    m^{2} &= F(r)\left(P^{t}\right)^{2}-\frac{\left(P^{r}\right)^{2}}{F(r)}-r^{2}\left(P^{\phi}\right)^{2},\label{eq:conservationm}\\
    S^{2} &= -\left(S^{tr}\right)^{2}+\frac{r^{2}}{F(r)}\left(S^{r\phi}\right)^{2}-r^{2}F(r)\left(S^{t\phi}\right)^{2}.\label{eq:conservationS}
\end{align}
On the other hand, the TSSC \cref{eq:Tulc_SSC} for our case reads
\begin{align}
    \frac{S^{tr}P^{r}}{F(r)}+r^{2}S^{t\phi}P^{\phi} &=0,\label{eq:ssc1}\\
    F(r) S^{tr}P^{t}+r^{2}S^{r\phi}P^{\phi} &=0,\label{eq:ssc2}\\
    F(r)S^{t\phi}P^{t}-\frac{S^{r\phi}P^{r}}{F(r)}&=0.\label{eq:ssc3}
\end{align}
\begin{widetext}
The four-momentum \cref{eq:cov_der_four_mom} for the metric (\ref{eq:metric1}) reduces to the following equations:
\begin{align}
    \dot{P^{t}} +\frac{F'(r)\dot{r}}{2F(r)}P^{t}+\frac{F'(r)}{2F(r)}P^{r} &= \frac{F''(r)\dot{r}}{2F(r)}S^{tr}+\frac{rF'(r)\dot{\phi}}{2}S^{t\phi},\label{eq:dotPt}
    \\
    \dot{P^{r}}+\frac{F(r)F'(r)}{2}P^{t}-\frac{F'(r)\dot{r}}{2F(r)}P^{r}-rF(r)\dot{\phi}P^{\phi}&=\frac{F(r)F''(r)}{2}S^{tr}+\frac{rF'(r)\dot{\phi}}{2}S^{r\phi},\label{eq:dotPr}
    \\
    \dot{P^{\phi}}+\frac{\dot{r}}{r}P^{\phi}+\frac{\dot{\phi}}{r}P^{r} &= \frac{F(r)F'(r)}{2r}S^{t\phi}-\frac{F'(r)\dot{r}}{2rF(r)}S^{r\phi},\label{eq:dotPphi}
\end{align}
and the equations of motion for the spin calculated from \cref{eq:cov_der_spin_tensor} turn out to be
\begin{align}
    \dot{S^{tr}}-rF(r)\dot{\phi}S^{t\phi}&=\dot{r}P^{t}-P^{r},\label{eq:dotStr}
    \\
    \dot{S^{t\phi}}+\frac{F'(r)}{2F(r)}S^{r\phi}+\dot{r}\left(\frac{F'(r)}{2F(r)}+\frac{1}{r}\right)S^{t\phi}+\frac{\dot{\phi}}{r}S^{tr} &= \dot{\phi}P^{t}-P^{\phi},\label{eq:dotStphi}\\
    \dot{S^{r\phi}}+\frac{F(r)F'(r)}{2}S^{t\phi}+\dot{r}\left(\frac{1}{r}-\frac{F'(r)}{2F(r)}\right)S^{r\phi}&=\dot{\phi}P^{r}-\dot{r}P^{\phi}.\label{eq:dotSrphi}
\end{align}
{where the over-dot sign ($\;\dot{}\;$) indicates the derivative with respect to coordinate time}. Using these 13 equations [i.e., \crefrange{eq:conservationE}{eq:dotSrphi}], one can completely determine the motion of a spinning particle in vicinity of a charged dyonic black hole in PFDM. Using \cref{eq:ssc1,eq:ssc2}, one can write
\begin{align}
\left[F(r)(P^{t})^{2}-\frac{(P^{r})^{2}}{F(r)}\right]\left(S^{tr}\right)^{2}=r^{4}\left[\frac{(S^{r\phi})^{2}}{F(r)}-F(r)(S^{t\phi})^{2}\right](P^{\phi})^{2}.\label{eq:relationbwssc1nssc2}    
\end{align}
\end{widetext}
Now, using conservation of mass and spin \cref{eq:conservationm,eq:conservationS} together with \cref{eq:relationbwssc1nssc2}, the expression for $S^{tr}$ comes out as
\begin{align}
    S^{tr}= \pm \frac{rSP^{\phi}}{m}.\label{eq:Str}
\end{align}
Using \cref{eq:conservationE,eq:conservationJ,eq:ssc2}, the relation between the conserved $E$ and $J$ is 
\begin{align}
    \left(E+\frac{S^{tr}F'(r)}{2}\right)S^{tr}+\left(J-r^{2}P^{\phi}\right)r P^{\phi}=0,\label{eq:new_relation_bw_E_J}
\end{align}
which can be solved explicitly for $P^{\phi}$ after substituting the value of $S^{tr}$ from \cref{eq:Str}. The expression for $P^{\phi}$  reads as
\begin{align}
    \frac{P^{\phi}}{m}=\frac{2r^2\left(\mathcal{J}- \mathcal{S}\mathcal{E}\right)}{2r^4+2 Q_{e+m}^2 S^2-\left(2M+\lambda\right)r S^2+r S^2\lambda\ln{\frac{r}{\abs{\lambda}}}},\label{eq:pphi}
\end{align}    
where $Q_{e+m}^2=Q^{2}_{e}+Q^{2}_{m}$, and $\mathcal{J}=J/m$, $\mathcal{S}=\mp S/m$ and $\mathcal{E}=E/m$ are the total angular momentum, spin and energy per unit mass, respectively. Here, $\mathcal{S}$ ({negative}) {positive} means that the spin of the particle is ({anti-}) parallel to $J$. {It is important to note here that total angular momentum per unit mass $\mathcal{J}$ is equal to the sum of spin angular momentum per unit mass $\mathcal{s}$ and orbital angular momentum per unit mass $L$.}  Similarly, using \cref{eq:conservationE,eq:Str,eq:pphi}, we find the time component of the four-momentum
\begin{align}
    \frac{F(r)P^{t}}{m}=\frac{2r^{4}\mathcal{E}+\left[Q^{2}_{e+m}-2Mr-r\lambda+r\lambda\ln{\frac{r}{\abs{\lambda}}}\right]\mathcal{J}\mathcal{S}}{2r^4+2 Q_{e+m}^2 S^2-\left(2M+\lambda\right)r S^2+r S^2\lambda\ln{\frac{r}{\abs{\lambda}}}},\label{eq:pt}
\end{align}
which further leads to the explicit expression of $P^{r}$ by using \cref{eq:pphi} and conservation of mass \cref{eq:conservationm}
\begin{widetext}
\begin{align}
    \frac{\left(P^r\right)^2}{m^2}=\left(
    \frac{2r^{4}\mathcal{E}+\left[Q^{2}_{e+m}-2Mr-r\lambda+r\lambda\ln{\frac{r}{\abs{\lambda}}}\right]\mathcal{J}\mathcal{S}}{2r^4+2 Q_{e+m}^2 S^2-\left(2M+\lambda\right)r S^2+r S^2\lambda\ln{\frac{r}{\abs{\lambda}}}}
    \right)^2-
      F(r)-\left(
    \frac{2r^3 \sqrt{F(r)}\left(\mathcal{J}- \mathcal{S}\mathcal{E}\right)}{2r^4+2 Q_{e+m}^2 S^2-\left(2M+\lambda\right)r S^2+r S^2\lambda\ln{\frac{r}{\abs{\lambda}}}}
    \right)^2.\label{eq:pr}
\end{align}
\end{widetext}
It is worth noting here that in the limits $\mathcal{S}\to 0$ and $\lambda\to 0$, the \cref{eq:pphi,eq:pt,eq:pr} reduce to the nonzero components of the four-momentum for the Schwarzschild black hole obtained in \cite{Armaza:2015eha}.

\begin{table*}
\begin{center}
\caption{Numerical values of radial component $r$ and bound on spin parameter $\mathcal{S}$ for which $\mathcal{A}$ is minimum. }\label{tab:table2}
\resizebox{\linewidth}{!}
{
\begin{tabular}{l l l l | l l | l l | l l}
 \hline \hline
 & &\multicolumn{2}{c|}{\;\;\;\;\;\;$Q_{e}=Q_{m}=0$} & \multicolumn{2}{c|}{$Q_{e}=0.4 \;\text {and}\; Q_{m}=0$} &\multicolumn{2}{c|}{$Q_{e}=0.5\; \text{and}\; Q_{m}=0$} &\multicolumn{2}{c}{$Q_{e}=0.6\; \text {and}\; Q_{m}=0$}\\
\cline{3-4}\cline{5-6}\cline{7-8}\cline{9-10}
& $\lambda/M$    & $r_{min}/M$ & $\mathcal{S}(\pm)$ & $r_{min}/M$ & $\mathcal{S}(\pm)$ & $r_{min}/M$ & $\mathcal{S}(\pm)$ & $r_{min}/M$ & $\mathcal{S}(\pm)$\\
\hline
& 0.0000001   & 3.00000 & $\pm$5.19615\;\;\;\; & 2.88924 & $\pm$5.05297 & 2.82287 & $\pm$4.96791 & 2.73693 & $\pm$4.85869\\
& 0.000001    & 2.99998 & $\pm$5.19611 & 2.88922 & $\pm$5.05294 & 2.82285 & $\pm$4.96787 & 2.73691 & $\pm$4.85865\\
& 0.00001     & 2.99982 & $\pm$5.19582 & 2.88905 & $\pm$5.05264 & 2.82268 & $\pm$4.96757 & 2.73673 & $\pm$4.85834\\
& 0.0001      & 2.99850 & $\pm$5.19347 & 2.88769 & $\pm$5.05023 & 2.82129 & $\pm$4.96512 & 2.73529 & $\pm$4.85584\\
& 0.001       & 2.98850 & $\pm$5.17536 & 2.87734 & $\pm$5.03167 & 2.81069 & $\pm$4.94627 & 2.72435 & $\pm$4.83656\\
& 0.01        & 2.91985 & $\pm$5.04869 & 2.80642 & $\pm$4.90209 & 2.73821 & $\pm$4.81474 & 2.64956 & $\pm$4.70223\\
& 0.1         & 2.56341 & $\pm$4.35581 & 2.43967 & $\pm$4.19613 & 2.36405 & $\pm$4.09965 & 2.26402 & $\pm$3.97353\\
& 0.2         & 2.35962 & $\pm$3.92406 & 2.23287 & $\pm$3.76087 & 2.15482 & $\pm$3.66158 & 2.05061 & $\pm$3.53078\\
& 0.3         & 2.24441 & $\pm$3.65108 & 2.11924 & $\pm$3.49040 & 2.04208 & $\pm$3.39255 & 1.93891 & $\pm$3.26346\\
& 0.4         & 2.18206 & $\pm$3.47439 & 2.06104 & $\pm$3.31964 & 1.98667 & $\pm$3.22564 & 1.88760 & $\pm$3.10198\\
& 0.5         & 2.15448 & $\pm$3.36190 & 2.03889 & $\pm$3.21475 & 1.96825 & $\pm$3.12578 & 1.87473 & $\pm$3.00935\\
& 0.6         & 2.15094 & $\pm$3.29430 & 2.04128 & $\pm$3.15535 & 1.97468 & $\pm$3.07181 & 1.88716 & $\pm$2.96313\\
\hline
& &\multicolumn{2}{c|}{$Q_{e}=0.2\;\text{and}\;Q_{m}=0.5$} & \multicolumn{2}{c|}{$Q_{e}=0.4 \;\text {and}\; Q_{m}=0.5$} &\multicolumn{2}{c|}{$Q_{e}=0.5\; \text{and}\; Q_{m}=0.5$} &\multicolumn{2}{c}{$Q_{e}=0.6\; \text {and}\; Q_{m}=0.5$}\\
\cline{3-4}\cline{5-6}\cline{7-8}\cline{9-10}
& $\lambda/M$    & $r_{min}/M$ & $\mathcal{S}(\pm)$ & $r_{min}/M$ & $\mathcal{S}(\pm)$ & $r_{min}/M$ & $\mathcal{S}(\pm)$ & $r_{min}/M$ & $\mathcal{S}(\pm)$\\
\hline
& 0.0000001   & 2.79228 & $\pm$4.92891\;\;\;\; & 2.69582 & $\pm$4.80686 & 2.61803 & $\pm$4.70959 & 2.51489 & $\pm$4.58250\\
& 0.000001    & 2.79226 & $\pm$4.92887 & 2.69580 & $\pm$4.80683 & 2.61801 & $\pm$4.70956 & 2.51486 & $\pm$4.58246\\
& 0.00001     & 2.79209 & $\pm$4.92857 & 2.69562 & $\pm$4.80651 & 2.61782 & $\pm$4.70924 & 2.51466 & $\pm$4.85213\\
& 0.0001      & 2.79068 & $\pm$4.92609 & 2.69416 & $\pm$4.80398 & 2.61631 & $\pm$4.70664 & 2.51307 & $\pm$4.57944\\
& 0.001       & 2.77997 & $\pm$4.90709 & 2.68303 & $\pm$4.78448 & 2.60479 & $\pm$4.68669 & 2.50095 & $\pm$4.55881\\
& 0.01        & 2.70670 & $\pm$4.77461 & 2.60701 & $\pm$4.64869 & 2.52616 & $\pm$4.54788 & 2.41816 & $\pm$4.41539\\
& 0.1         & 2.32875 & $\pm$4.05493 & 2.21513 & $\pm$3.91261 & 2.12024 & $\pm$3.79596 & 1.98782 & $\pm$3.63747\\
& 0.2         & 2.11818 & $\pm$3.61535 & 1.99918 & $\pm$3.46709 & 1.89807 & $\pm$3.34389 & 1.75277 & $\pm$3.17281\\
& 0.3         & 2.00583 & $\pm$3.34695 & 1.88791 & $\pm$3.20050 & 1.78739 & $\pm$3.07846 & 1.64207 & $\pm$2.90815\\
& 0.4         & 1.95181 & $\pm$3.18191 & 1.83881 & $\pm$3.04186 & 1.74314 & $\pm$2.92573 & 1.60636 & $\pm$2.76492\\
& 0.5         & 1.93526 & $\pm$3.08452 & 1.82898 & $\pm$2.95303 & 1.74001 & $\pm$2.84496 & 1.61522 & $\pm$2.69736\\
& 0.6         & 1.94372 & $\pm$3.03320 & 1.84466 & $\pm$2.91089 & 1.76277 & $\pm$2.81134 & 1.65015 & $\pm$2.67738\\
 \hline \hline
\end{tabular}
}
\end{center}
\end{table*}

Next, we are interested in the explicit form of the radial and azimuthal components of coordinate velocity (i.e., $\dot{r}$ and $\dot{\phi}$){, since they will help us later to confine the motion of the spin particle to the subluminal zone, which is the region where the particle's four-velocity is timelike}. With this aim in mind, we need to first find the unknown components of the spin tensor (i.e., $S^{t\phi}$ and $S^{r\phi}$) in terms of known four-momentum components. By using the first [\cref{eq:ssc1}] and second [\cref{eq:ssc2}] TSSC together with \cref{eq:Str} it is easy to show that the components $S^{t\phi}$ and $S^{r\phi}$ of the spin tensor can be written as
\begin{align}
    S^{t\phi}&= \frac{\mathcal{S} P^{r}}{rF(r)}\label{eq:stphi}\\
    \text{and}\hspace{0.5cm}& \nonumber\\
    S^{r\phi}&= \frac{\mathcal{S}F(r)P^{t}}{r}.\label{eq:srphi}
\end{align}
To determine $\dot{r}$ first, we multiply the factor $\pm(\mathcal{S}r)$ to the TSSC \cref{eq:ssc3} and subtract \cref{eq:dotStr} together with the help of \cref{eq:stphi,eq:srphi}, we get
\begin{align}
    -\dot{S^{tr}}+\mathcal{S}P^{r}\dot{\phi}=-\dot{r}P^{t}+P^{r}
\end{align}
This, when combined with \cref{eq:dotPphi,eq:Str} along with \cref{eq:stphi,eq:srphi}, allows us to find the precise symbolic expression as
\begin{align}
    \dot{r}{\equiv\frac{dr}{dt}}=\frac{P^{r}}{P^{t}},\label{eq:dotr1}
\end{align}
and now, to calculate $\dot{\phi}$, we utilize \cref{eq:Str,eq:pphi,eq:stphi,eq:dotr1} in \cref{eq:dotStr}. With a little algebra, we obtain
\begin{align}
    \dot{\phi}{\equiv\frac{d\phi}{dt}}=\left[\frac{2-F''(r)\mathcal{S}^2}{2\left(1-\frac{F'(r)\mathcal{S}^2}{2r}\right)}\right]\frac{P^{\phi}}{P^t}.\label{eq:dotphi}
\end{align}
It is also observed that the spatial component of the spin perpendicular to the $\theta=\pi/2$ plane is
\begin{align}
    S_{z}&=r S^{r\phi},\nonumber\\
         &=m\mathcal{S}\left(\frac{2r^{4}\mathcal{E}+\left[Q^{2}_{e+m}-2Mr-r\lambda+r\lambda\ln{\frac{r}{\abs{\lambda}}}\right]\mathcal{J}\mathcal{S}}{2r^4+2 Q_{e+m}^2 S^2-\left(2M+\lambda\right)r S^2+r S^2\lambda\ln{\frac{r}{\abs{\lambda}}}}\right),\label{eq:sz}
\end{align}
which in the asymptotic limit (i.e., $r\to\infty$) becomes $S_{z}/m=\mathcal{S}\mathcal{E}$. {It is also worth mentioning that, as demonstrated in \cite{Zalaquett:2014eia,Armaza:2015eha}, one may simply bypass the four-momentum and four-velocity relation \cref{eq:rel_bw_u_n_V_2} and work with coordinate velocities solely in the case of static spherically symmetric spacetimes. As a result, we directly found the necessary coordinate velocities in \cref{eq:dotr1,eq:dotphi} (i.e. $\dot{r}$ and $\dot{\phi}$, respectively) instead of the four-velocity of the spinning particle.}

\begin{figure*}
\begin{tabular}{c c}
  \includegraphics[scale=0.6]{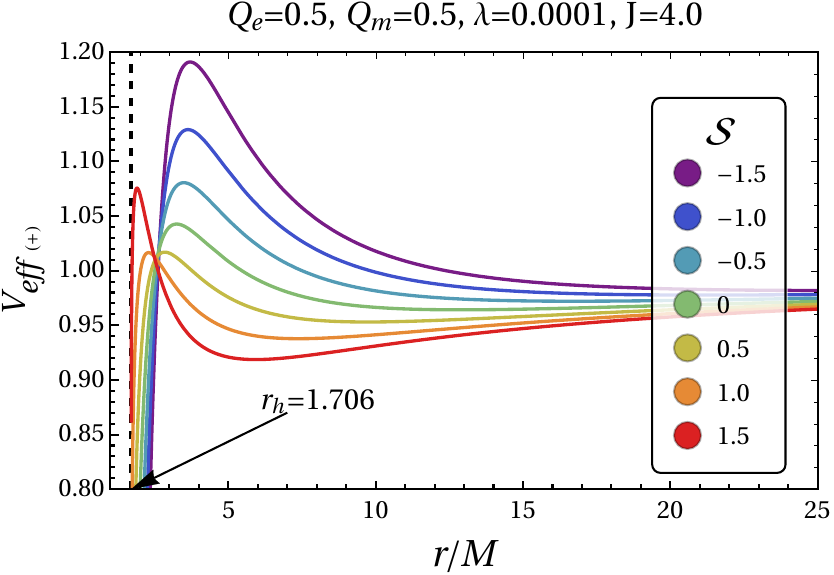}\hspace{0cm}
  &  \includegraphics[scale=0.6]{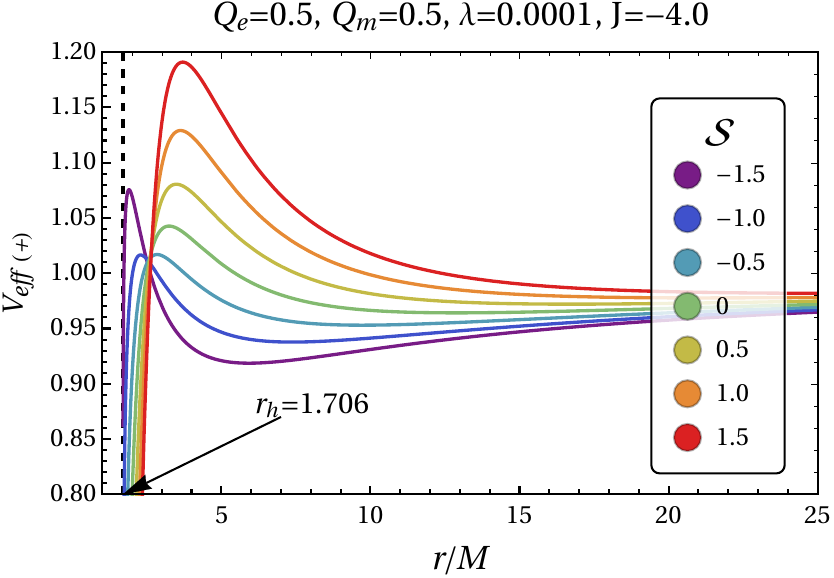}
\end{tabular}
	\caption{\label{fig:Veff1} 
Effective potential $V_{eff(+)}$ as a function of the spatial distance $r/M$. 
The {left panel} corresponds to direct trajectories, whereas the {right panel} refers to the retrograde trajectories {for the different values of parameter $\mathcal{S}$, which varies from -1.5 to 1.5 in the steps of 0.5}.
}
\end{figure*}
\begin{figure*}
\begin{tabular}{c c}
 \includegraphics[scale=0.6]{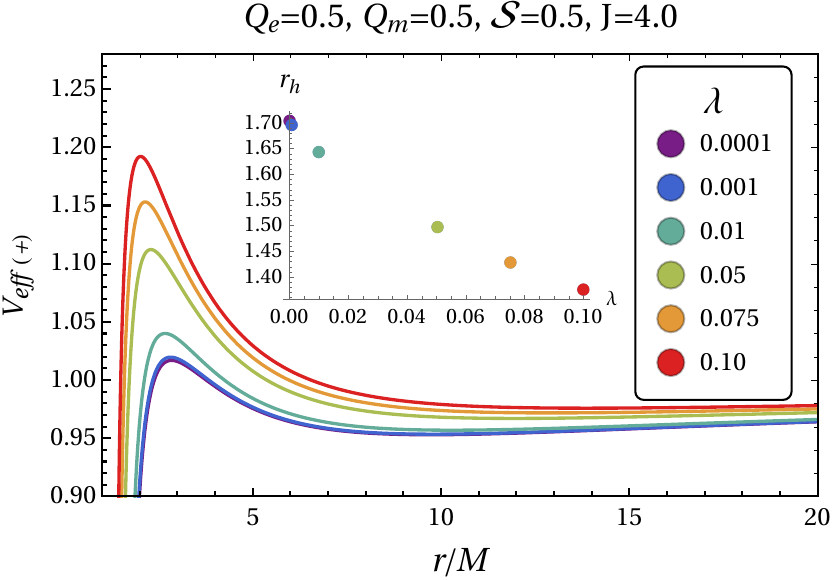}\hspace{0cm}
  &  \includegraphics[scale=0.6]{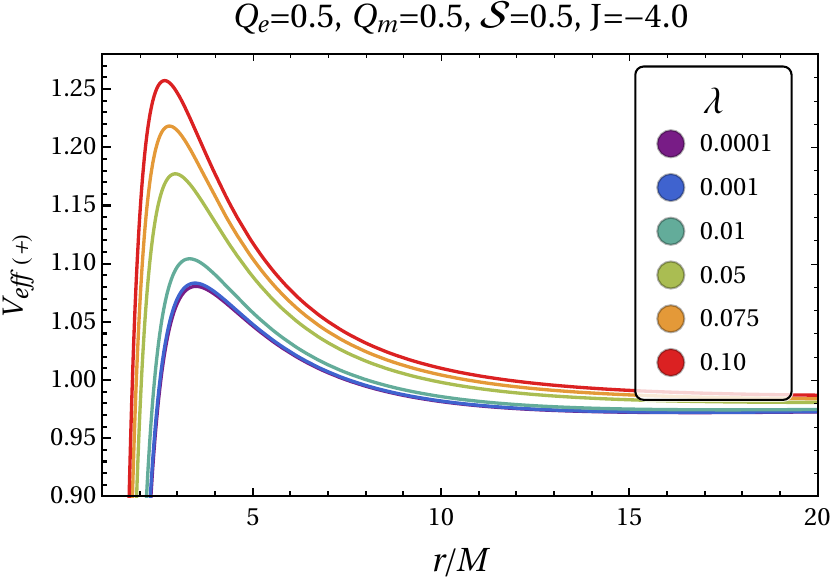} 
 \end{tabular}
	\caption{\label{fig:Veff2} 
	Effective potential $V_{eff(+)}$ as a function of the spatial distance $r/M$. The {left panel} corresponds to direct trajectories, whereas the {right panel} is for retrograde trajectories. 
 The inset graphic in the {left panel} depicts the evolution of the event horizon $r_{h}$ as the PDFM parameter $\lambda$ increases.}
\end{figure*}
\begin{figure*}
\begin{tabular}{c c}
 \includegraphics[scale=0.6]{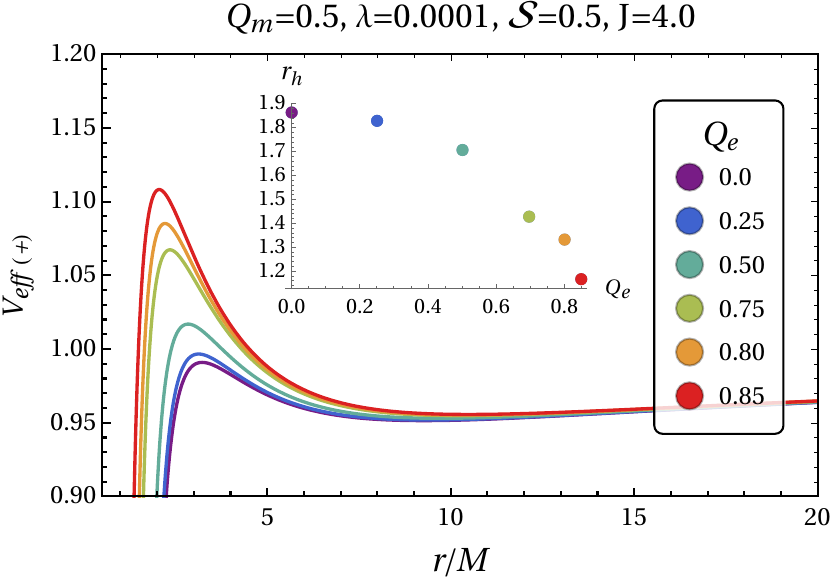}\hspace{0cm}
  &  \includegraphics[scale=0.6]{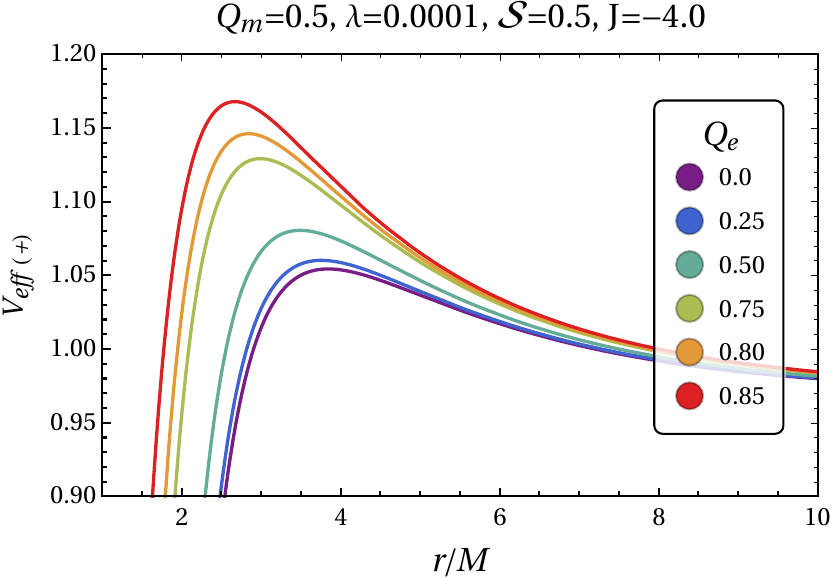} 
  \\
  \includegraphics[scale=0.6]{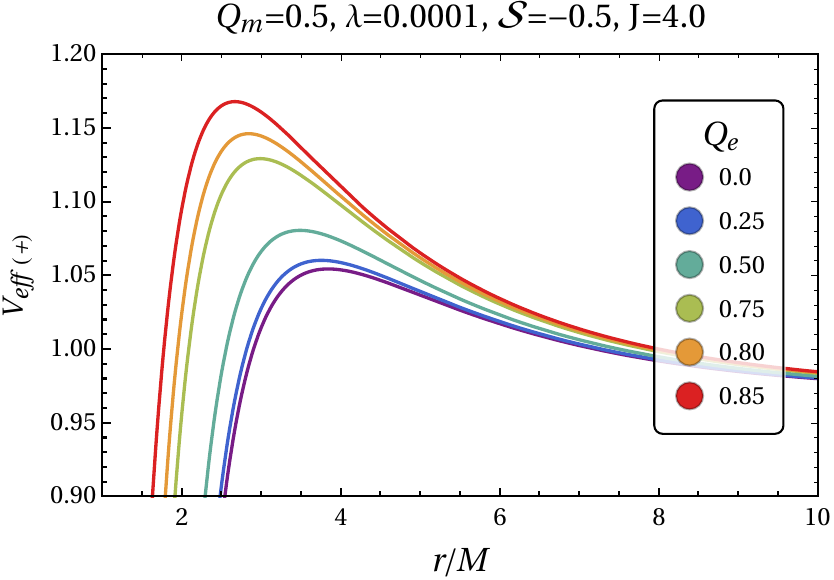}\hspace{0cm}
  &  \includegraphics[scale=0.6]{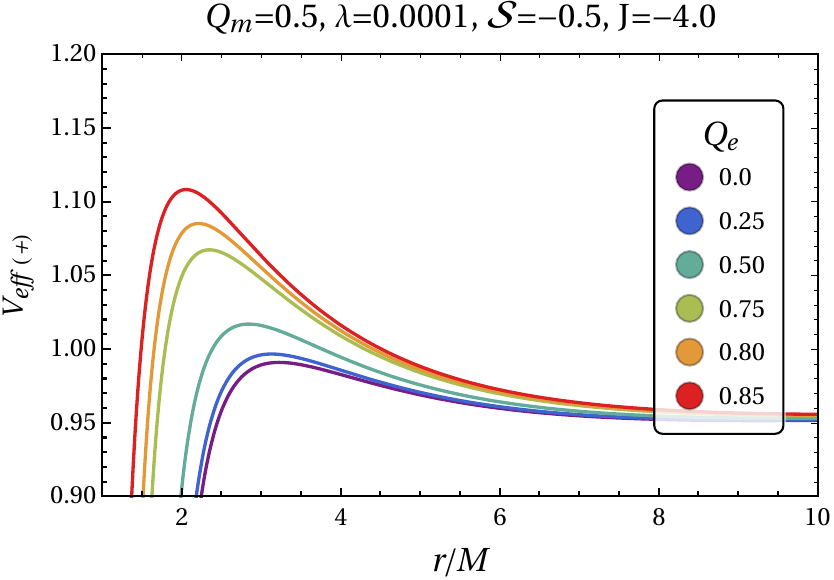}  
\end{tabular}
	\caption{\label{fig:Veff3} 
	Effective potential $V_{eff(+)}$ as a function of the spatial distance $r/M$. The {left column} corresponds to direct trajectories, whereas the {right column} is for retrograde trajectories. 
 The inset graphic in the {left panel} of the top row depicts the evolution of the event horizon $r_{h}$ as the parameter $Q_{e}$ increases. Here, in the {top row} the spin per unit mass $\mathcal{S}$ is fixed to value $0.5$, and in the {bottom row} the spin per unit mass $\mathcal{S}$ is fixed to value $-0.5$, respectively.}
\end{figure*}

\subsection{Effective potential}
\label{effective_pot}
It is well known from Newtonian mechanics that the equations of motion can be solved in terms of the radial coordinate \cite{LandauandLifshitz} to study the motion of a test particle in a central force field, and that in general relativity, the ``effective potential" method is widely used to study the dynamics of a test particle in the black hole background.

Because the radial velocity $u^r$ is parallel to the radial component $P^{r}$ of the four-momentum $P^{\mu}$ \cite{Jefremov:2015gza}, the effective potential $V_{eff}$ of the spinning particle moving in the background of a charged dyonic black hole in PFDM can be computed using the $P^{r}$ \cref{eq:pr}. Since this equation is quadratic in $\mathcal{E}$, factorizing it separates the energy component from the radial component, giving us
\begin{align}
    \frac{(P^{r})^2}{m^2}=\mathcal{A}\mathcal{B}^{-2}\left[\mathcal{E}-V_{eff(+)}(r)\right]\left[\mathcal{E}-V_{eff(-)}(r)\right],\label{eq:Pr2_Veff}
\end{align}
where, $\mathcal{A,B}$ and the $V_{eff(\pm)}$ are
\begin{align}
    V_{eff(\pm)}(r)\equiv V_{eff(\pm)}=\frac{\mathcal{C}\pm \sqrt{\mathcal{D}}}{2\mathcal{A}},\label{eq:Veff}
\end{align}
were defined, with
\begin{align}
    \mathcal{A}&\equiv 1+\frac{\mathcal{S}^{2}\mathcal{X}}{r^{4}},\;\;\;\;
    \mathcal{B}\equiv 1+\frac{\mathcal{S}^{2}\mathcal{Y}}{2r^{4}},\\
    \mathcal{C}&\equiv -\frac{\mathcal{J}\mathcal{S}}{r^4}\left(\mathcal{Y}-2\mathcal{X}\right),\\
    \mathcal{D}&\equiv \left(\frac{\mathcal{JS}}{r^2}\right)^{2}\left(\frac{\mathcal{Y}-2\mathcal{X}}{r^{2}}\right)^{2}-\mathcal{A}\Bigg[4\mathcal{J}^{2}\mathcal{X}\\\nonumber
    &+\left(\frac{\mathcal{JSY}}{r^{2}}\right)^{2}+2\mathcal{X}r^{2}\mathcal{B}^{2}\Bigg],\\
    \mathcal{X}&\equiv -\lambda  r \ln {\frac{r}{\left| \lambda \right| }}+2 M r-Q_{e+m}^2-r^2,\\
    \mathcal{Y}&\equiv \lambda  r \ln{ \frac{r}{\left| \lambda \right| }}-2 M r+2 Q_{e+m}^2-\lambda  r.
\end{align}
The first-look analysis of \cref{eq:Pr2_Veff} confirms that $P^{r}$ is real, and therefore $u^{r}$ if and only if the motion of the spinning particle is constrained in such a way that $\mathcal{E}>V_{eff(+)}$ or $\mathcal{E}<V_{eff(-)}$, {as mentioned in previous section as well together with the condition with $\mathcal{A}>0$, which is an extra condition coming solely because of the spin-orbit coupling}. In \cref{tab:table2}, we discover the values of the radial parameter (i.e. $r_{min}$) for various combinations of $\lambda, Q_{e}$ and $Q_{m}$, as well as the corresponding range of spin parameter per unit mass $\mathcal{S}$ for which the condition $\mathcal{A}>0$ is fulfilled.
For a consistency check, we demonstrated in \cref{tab:table2} that when $Q_{e}$ and $Q_{m}$ vanish and the PFDM parameter is extremely tiny (i.e., $\lambda=0.0000001$), the exact Schwarzschild limiting values (i.e., $r_{min}=3$ and $-5.19615<\mathcal{S}<5.19615$) as found in \cite{Armaza:2015eha} are recovered.
Furthermore, in the limit $Q_{m}\to0$ and $\lambda\to0$, the $V_{eff}$ obtained in \cref{eq:Veff} reduces to that of the Reissner-Nordstr\"{o}m black hole (BH). However, if $Q_{e}\to0$ in addition to $Q_{m}\to0$ and $\lambda\to0$, $V_{eff(+)}$ matches that of the Schwarzschild BH \cite{Armaza:2015eha}. 
The effective potential $V_{eff(+)}$ for both direct (i.e. $\mathcal{L}=\mathcal{J}-\mathcal{S}>0$) as well as retrograde (i.e. $\mathcal{L}=\mathcal{J}-\mathcal{S}<0$) trajectories of a spinning particle is plotted in \cref{fig:Veff1,fig:Veff2,fig:Veff3}, for various combinations of $Q_{e}, Q_{m}, \lambda$ and $\mathcal{S}$. A brief summary of the results obtained after analyzing the plots in \cref{fig:Veff1,fig:Veff2,fig:Veff3} of $V_{eff(+)}$ are as follows:
\begin{enumerate}[label=(\roman*)]
        \item In \cref{fig:Veff1}, 
        for the direct trajectories, we observe that when the parameter $\mathcal{S}$ grows, the maximum of $V_{eff(+)}$ first drops and then climbs again, as well as moves closer to the BH's event horizon $r_{h}$ (see {left panel}). In contrast, the maximum of $V_{eff(+)}$ goes {away} from the $r_{h}$ as the parameter $\mathcal{S}$ increases for retrograde trajectories (see {right panel}).
        \item In \cref{fig:Veff2}, we fix the parameters $Q_{e}, Q_{m}$ and $\mathcal{S}$ while varying the PFDM parameter $\lambda$. The maximum of $V_{eff(+)}$ grows as the parameter $\lambda$ increases for the direct orbits, as shown in the {left panel}. The variation of $r_{h}$ for the corresponding values of $\lambda$ is shown in the inset plot; we find that increasing the parameter $\lambda$ decreases the size of the event horizon $r_{h}$.
        A similar behavior is observed for the retrograde case of the $V_{eff (+)}$ in the {right panel} when the parameter $\lambda$ increases; the only difference is that the maximum of $V eff (+)$ is at higher values in comparison to the direct case for the equivalent values of $\lambda$.
        \item Similarly, in \cref{fig:Veff3}, the behavior of $V_{eff(+)}$ for both {direct (see left column) as well as retrograde trajectories (see right column) is plotted} as a function of $r/M$ is shown for various combinations of the parameters $Q_{m}, \lambda$ and $\mathcal{S}$ while varying the electric charge parameter $Q_{e}$. {The maximum of $V{eff(+)}$ is shown to grow when the parameter $Q_{e}$ rises for both direct and retrograde trajectories.} Additionally, it is seen from the inset plot in the {top row, left panel} that the event horizon $r_{h}$ decreases with an increase in {parameter} $Q_{e}$. It is also seen that the maximum of $V_{eff}(+)$ occurs at a higher value for the retrograde trajectories when $\mathcal{S}>0$ (see {top row} of \cref{fig:Veff3}), while for the case when $\mathcal{S}<0$, it occurs for the direct trajectories (see {bottom row} of \cref{fig:Veff3}).
    \end{enumerate}

\begin{figure*}
\begin{tabular}{c c}
  \includegraphics[scale=0.5]{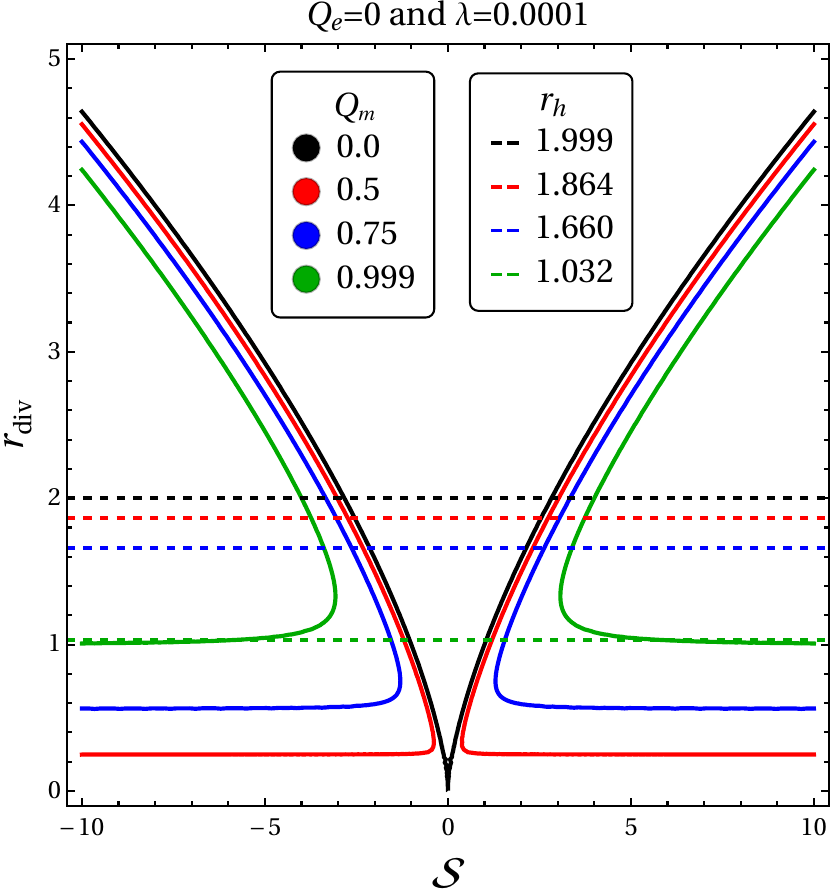}\hspace{0cm}
  &  \includegraphics[scale=0.5]{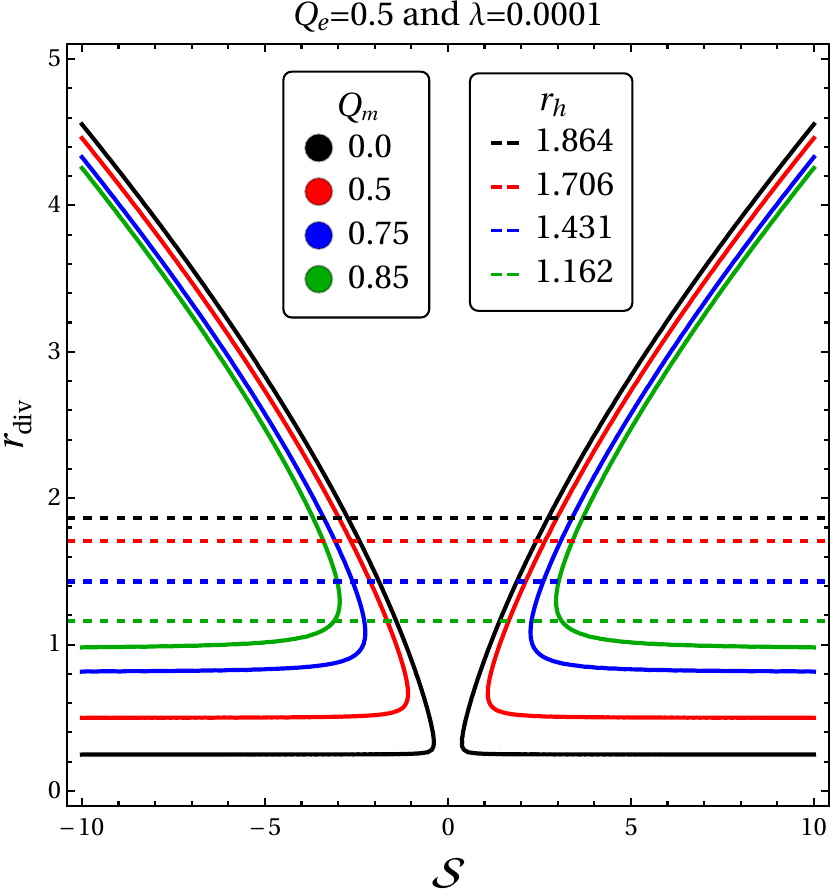}
\end{tabular}
	\caption{\label{fig:div_r_for_diff_Q} 
	Divergence radius $r_{div}$ as a function of spin parameter $\mathcal{S}$. In both panels {left} and {right}, we fix the parameter $\lambda$ and vary parameter $Q_{m}$. However, the parameter $Q_{e}=0$ and $0.5$ for the {left panel} and the {right panel}, respectively. Here, the horizontal dashed lines represent the location of event horizon $r_{h}$ for the corresponding combination of $Q_{e}, Q_{m}$ and $\lambda$. The mass parameter $M$ of a charged dyonic black hole immersed in PFDM is set to unity.}
\end{figure*}
\begin{figure*}
\begin{tabular}{c c c}
  \includegraphics[scale=0.4]{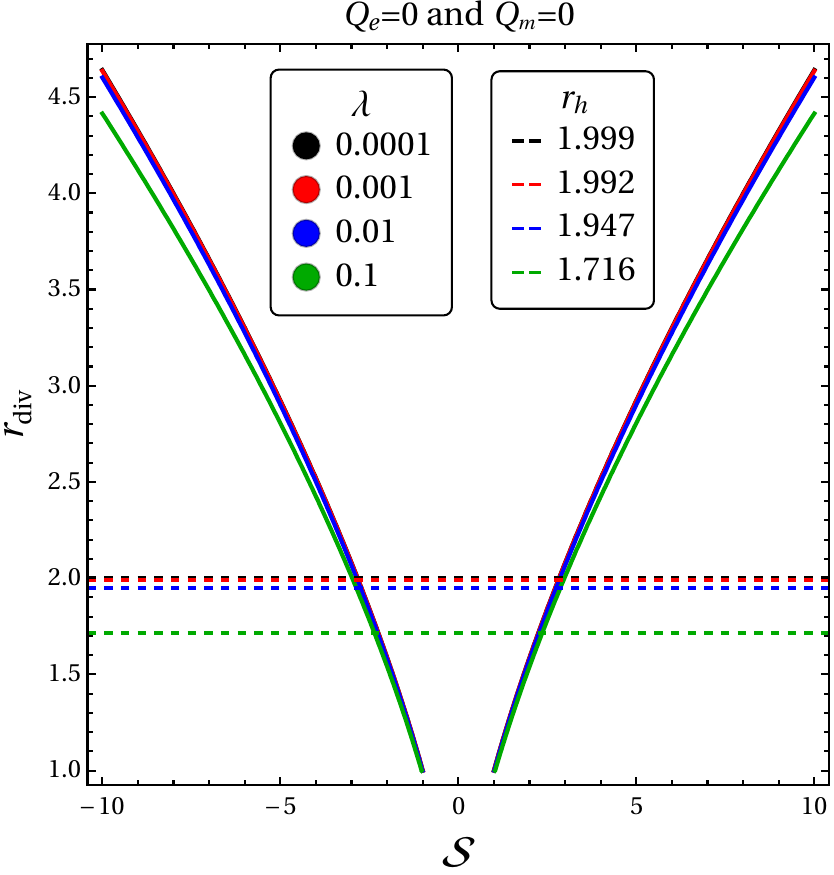}\hspace{0cm}
  &  \includegraphics[scale=0.4]{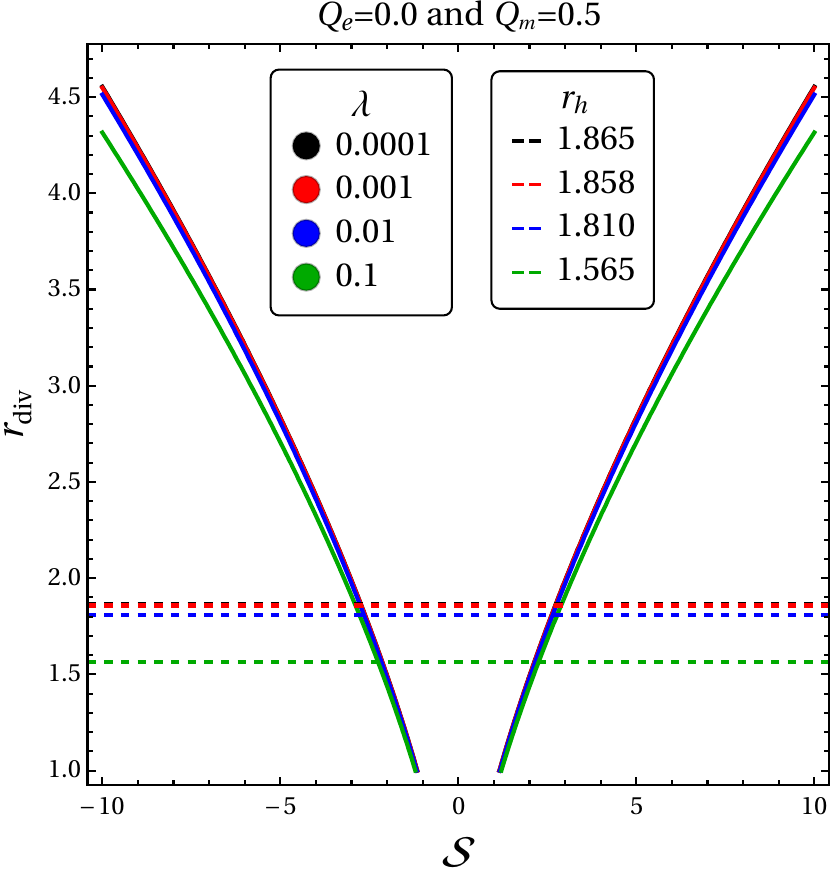}
 & \includegraphics[scale=0.4]{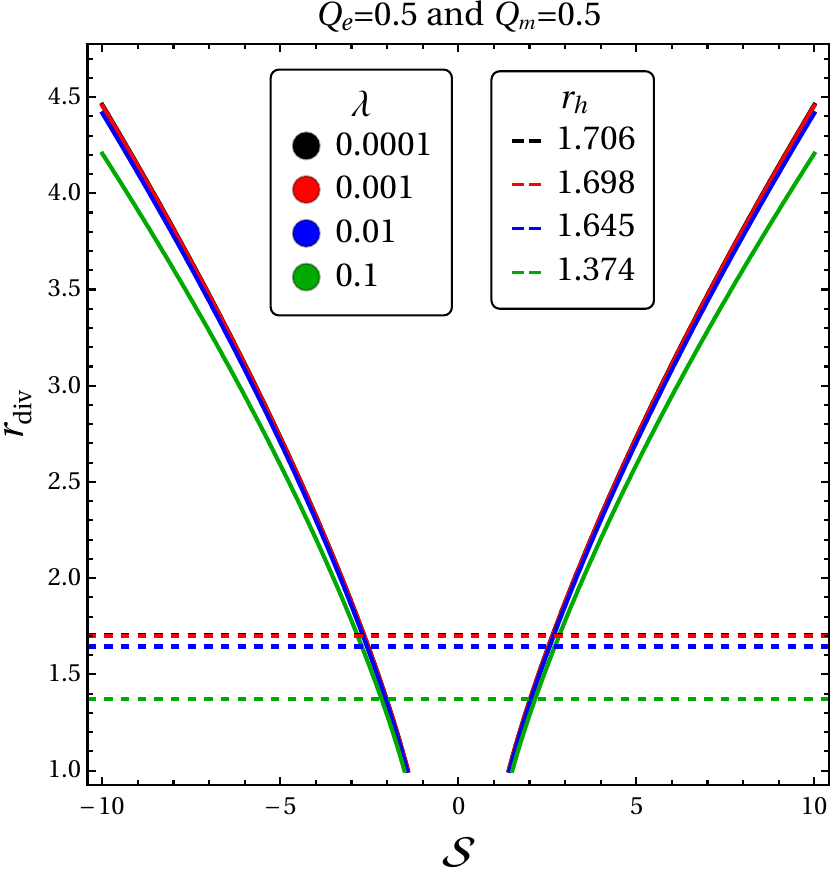}
\end{tabular}
	\caption{\label{fig:div_r_for_diff_lam} 
	Divergence radius $r_{div}$ as a function of spin $\mathcal{S}$ for different combinations of $Q_{e}+Q_{m}$ and  various parameters $\lambda$. The sum $Q_{e}+Q_{m}$ increases moving from left to right along the row. Here, also the mass parameter $M$ of a charged dyonic black hole immersed in PFDM is set to unity.}
\end{figure*}

Now, by analyzing \cref{eq:Pr2_Veff} further, one finds that the divergence of $(P^{r}/m)^2$ when $\mathcal{B}=0$ means that it gives the location 
$r_{div}$. 
As it is not easy to find the analytical value $r_{div}$, we thus numerically analyze it, and the results are presented in \cref{fig:div_r_for_diff_Q,fig:div_r_for_diff_lam}. In \cref{fig:div_r_for_diff_Q}, we plot $r_{div}$ as a function of the parameter $\mathcal{S}$ for the fixed values of $Q_{e}=0\; \text{and}\;0.5$, and $\lambda=0.0001$ and vary the magnetic charge parameter $Q_{m}$. We find that $r_{div}$ increases as the parameter $\pm\mathcal{S}$ increases. Similarly, in \cref{fig:div_r_for_diff_lam}, we plot the $r_{div}$ as a function of the parameter $\mathcal{S}$ but here we choose different combinations of $Q_{e}$ and $Q_{m}$, and vary the $\lambda$. A similar behavior $r_{div}$ is observed with the parameter $\pm\mathcal{S}$ as in \cref{fig:div_r_for_diff_Q}. 
{It is vital to notice that the parameters $Q_{e}$ and $Q_{m}$ were chosen in such a way that a black hole horizon should exist. However, the PFDM parameter $\lambda$ is selected in such a way that it must satisfy the requirement $\lambda \ll M$, as stated in \cref{Sec:metric}. Whereas the parameter $\mathcal{S}$ is selected in such a manner that it must be smaller than $M$ (known as the M\o ller limit \cite{Moller:1949, PhysRevD.6.406}), for the cases like the intermediate mass ratio inspirals, there is no such constraint on the choices of parameter $\mathcal{S}$. As a result, we picked both scenarios where the parameter $\mathcal{S}<M$ and $\mathcal{S}> M$.}

\begin{figure*}
\begin{tabular}{c c}
 \includegraphics[scale=0.5]{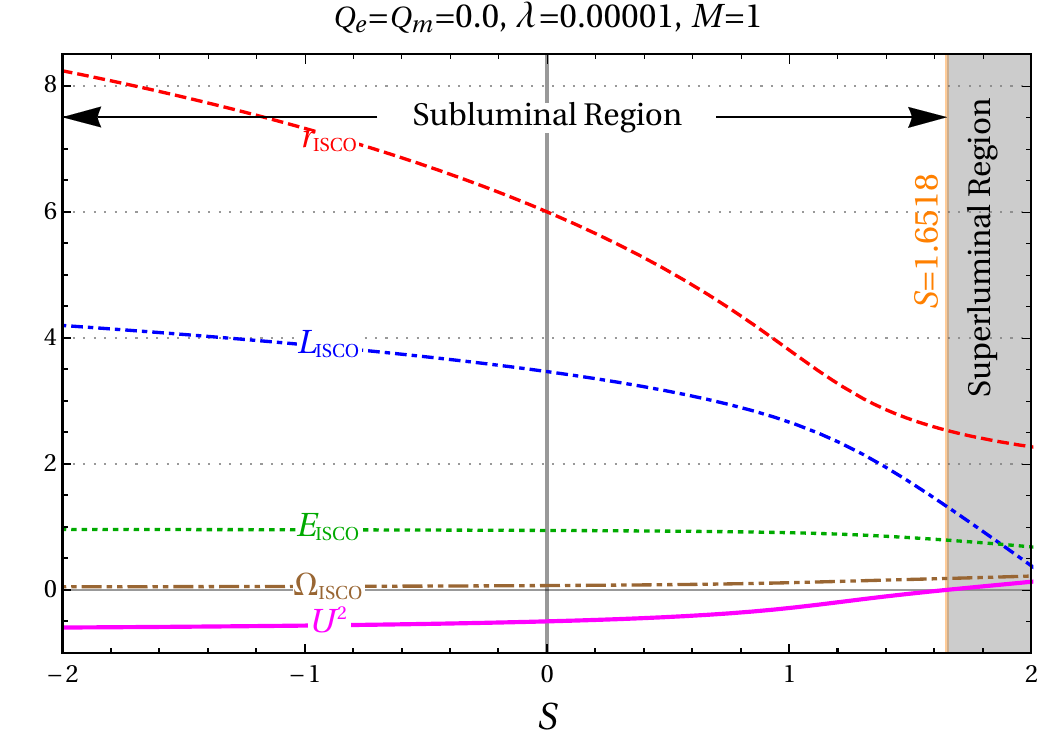}\hspace{0cm}
  &  \includegraphics[scale=0.5]{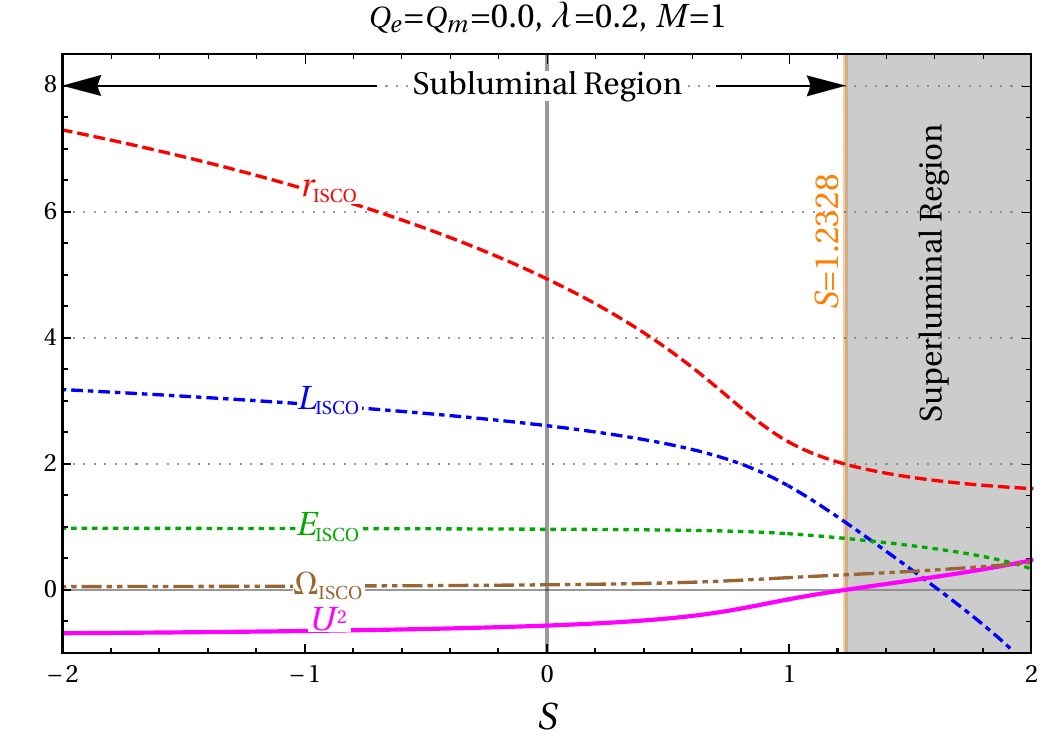}
  \\
  \includegraphics[scale=0.5]{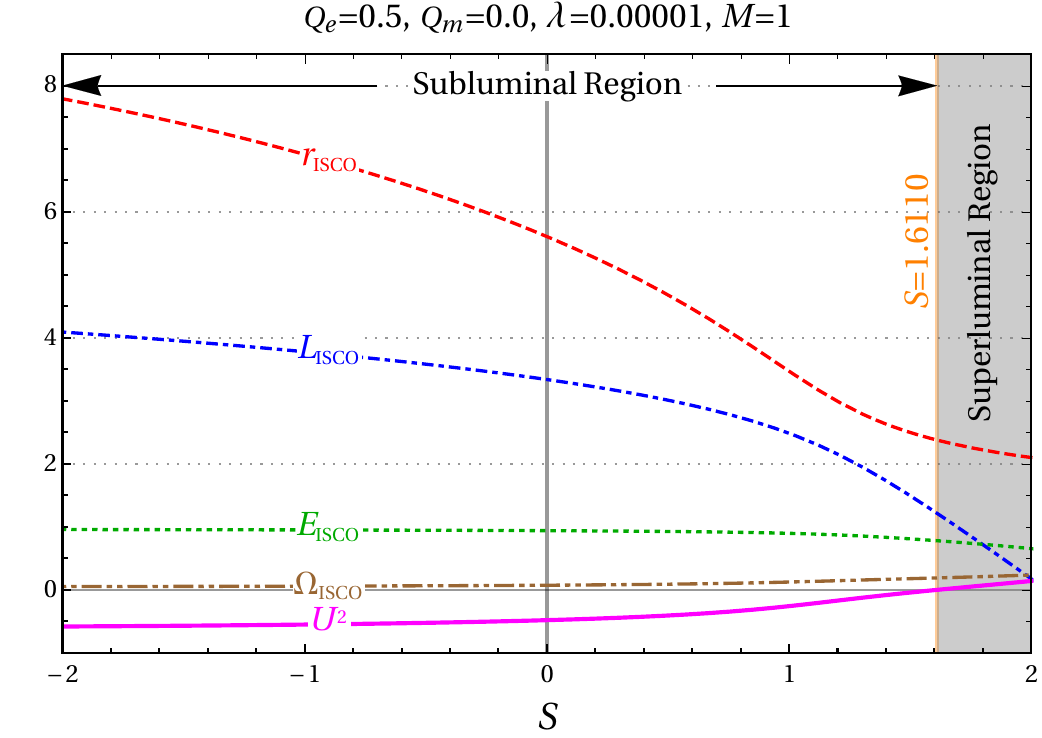}\hspace{0cm}
  &  \includegraphics[scale=0.5]{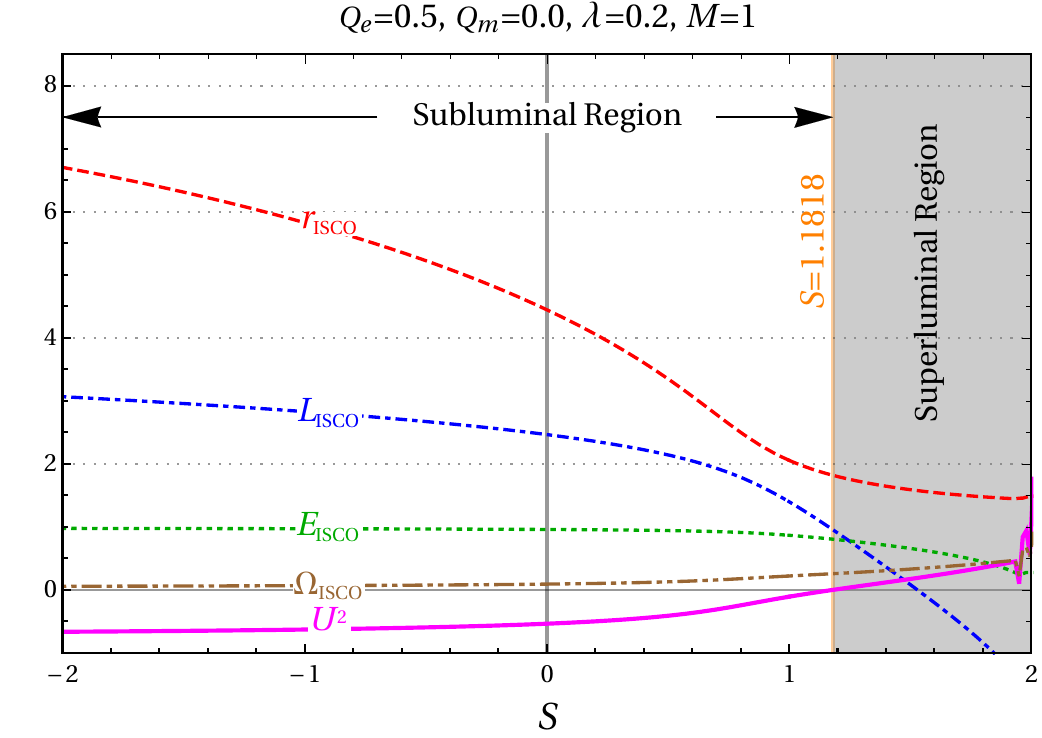}
  \\
  \includegraphics[scale=0.5]{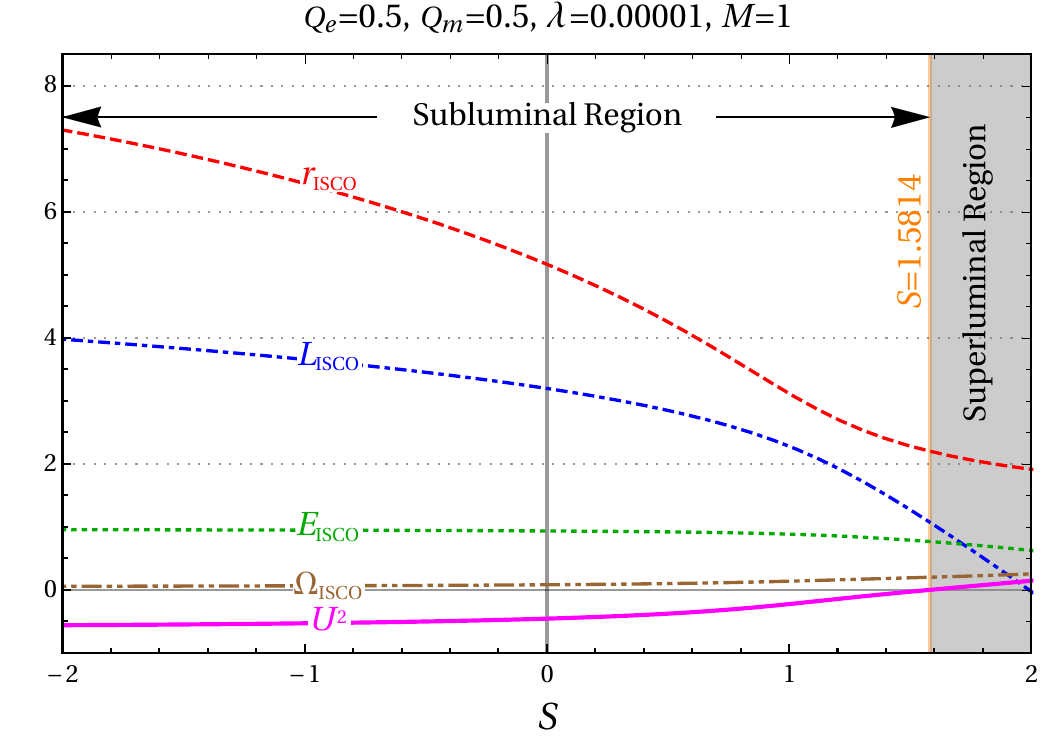}\hspace{0cm}
  &  \includegraphics[scale=0.5]{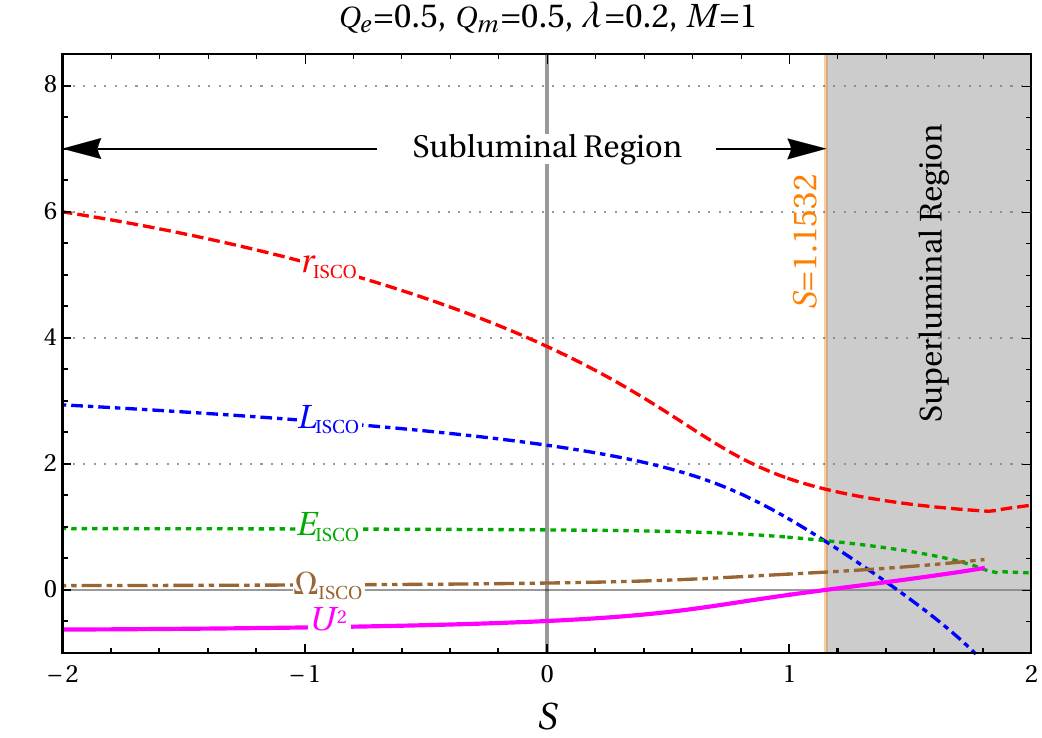}
\end{tabular}
	\caption{\label{fig:ISCO} 
Behavior of ISCO parameters $r_{ISCO}, L_{ISCO}, E_{ISCO}$ and $\Omega_{ISCO}$ as well as four-velocity square $U^{2}$ as a function of the spin per unit mass $\mathcal{S}$ for the various combinations of $Q_{e}, Q_{m}$ and $\lambda$. The PFDM parameter $\lambda$ grows as we move from left to right along each row. However, as we move down along each of the columns, the sum $Q_{e}+Q{m}$ increases. The black hole mass parameter $M$ is fixed to unity.}
\end{figure*}
\subsection{ISCO of a spinning particle moving in the background of a charged dyonic black hole immersed in PFDM}\label{sec:ISCO_spinning}

In this subsection, our primary interest is to study the behavior of the parameters $r_{ISCO}, L_{ISCO}, E_{ISCO},\;\text{and}\;\Omega_{ISCO}$ of a spinning particle numerically, and also, due to the complexity of the equations of motion, it is more convenient to define the effective potential similar to \cite{Jefremov:2015gza}. Hence, we redefine the effective potential using \cref{eq:pr} as
\begin{align}
    W_{eff}\equiv\left(\frac{P^{r}}{m}\right)^{2}.\label{eq:Weff}
\end{align}

Now, in order to find the ISCOs of the spinning particle, the analogy of \cref{eq:dotr11,eq:ddotr1} implies that we need to solve the following two equations simultaneously:
\begin{align}
   &\frac{dr}{d\tau}=0 \implies W_{eff}=0 \label{eq:ISCO1}\\
    \text{and}\;\;\; &\frac{d^{2}r}{d\tau^2}=0 \implies \frac{dW_{eff}}{dr}=0, \label{eq:ISCO2}
\end{align}
together with the condition [coming from the analogy of \cref{eq:ddVeff}] 
\begin{align}
    \frac{d^{2}W_{eff}}{dr^2}=0,\label{eq:ISCO3}
\end{align}
which gives the location of a flex point at which the maximum and minimum of $W_{eff}$ coincide.
Thus, \cref{eq:ISCO1,eq:ISCO2,eq:ISCO3} form a system of equations for determining the ISCO parameters $r_{ISCO}, L_{ISCO},\;\text{and}; E_{ISCO}$. The explicit forms \cref{eq:ISCO1,eq:ISCO2,eq:ISCO3} are shown in  \cref{Sec:Appendix_B} as \cref{eq:ISCO11,eq:ISCO21,eq:ISCO31}. 
The evolution of $r_{ISCO}, L_{ISCO},\;\text{and}; E_{ISCO}$, which are thus dependent on the black hole parameters  $Q_{e}$, $Q_{m}$, $M$, {PFDM parameter $\lambda$}, and the particle's spin $S$, is thus analyzed {in \cref{fig:ISCO}} using these three equations, which constitute a closed system.
It is worth mentioning here that for the spinning particle we use the total angular momentum $J$ which is the sum of orbital momentum $L$ and spin $S$ {(i.e. $J=L+S$)} \cite{Saijo:1998mn}\footnote{We work with the z component of total angular momentum, orbital angular momentum, and spin (i.e.,  $J_{z}, L_{z}$ and $S_{z}$) throughout the paper, but we drop the subscript z everywhere for simplicity only.}

In addition, for $r_{ISCO}, L_{ISCO}\;\text{and}; E_{ISCO}$ there is one more important quantity related to the circular motion of the particle that is its orbital angular velocity $\Omega$, also known as orbital angular frequency or Keplerian angular frequency. 
This Keplerian frequency $\Omega$ is identified via the relation
\begin{align}
    \Omega\equiv\dot{\phi}.\label{eq:Omega}
\end{align}
To find the $\Omega_{ISCO}$ for a spinning particle, we use the analogy of a geodesic particle, and hence substitute the values of $r_{ISCO}, L_{ISCO},\;\text{and}; E_{ISCO}$ in \cref{eq:Omega} found via \cref{eq:ISCO1,eq:ISCO2,eq:ISCO3}.
{We analyze the behavior of $\Omega$ as a function of $\mathcal{S}$ at the ISCO, as seen by an observer at infinity in \cref{fig:ISCO} together with other ISCO parameters.}
{Further,} while studying the motion of a spinning particle in a curved spacetime { it is important to take into consideration superluminal constraint.} 
As we know from the literature \cite{Suzuki:1997by,Jefremov:2015gza}, for a spinning particle moving in curved spacetime, its four-momentum $P^{\mu}$ and four-velocity $u^{\mu}$ are no longer parallel to each other. Hence, four-velocity can be timelike (i.e., $u_{\mu}u^{\mu}<0$) as well as spacelike (i.e., $u_{\mu}u^{\mu}>0$). The timelike (physical) four-velocity {of a spinning particle} is known as the subluminal, whereas the spacelike (unphysical) four-velocity {of a spinning particle} is known as superluminal.  Therefore, to ensure that the motion of the spinning particle is subluminal (physical), we impose a constraint
\begin{align}
    U^{2}=\frac{u_{\mu}u^{\mu}}{(u^{t})^2}=-F(r)+\frac{1}{F(r)}\left(\dot{r}\right)^{2}+r^{2}\left(\dot{\phi}\right)^{2}<0 \label{eq:U2}
\end{align}
known as the superluminal constraint. The explicit form of the $U^{2}$ is given in \cref{Sec:Appendix_B} as \cref{eq:U2_1}.
{The behavior of $U^{2}$ as a function of parameter $\mathcal{S}$ in addition to the behavior of parameters $r_{ISCO}, L_{ISCO}, E_{ISCO}$, and $\Omega_{ISCO}$ is shown in \cref{fig:ISCO}, for various values of black hole parameters $\lambda$, $Q_{e}$ and $Q_{m}$.}  

\section{Summary and Conclusions\label{Sec:conclusion}}
{In this paper, we have investigated the motion of two different classes of test particles: (\textit{i}) a massive test particle with charge, and (\textit{ii}) a massive particle with spin $S$, moving in the space-time of a charged dyonic black hole immersed in PFDM. To investigate the aforementioned, we first described the line element of the black hole using \cref{eq:metric1,eq:F}, and {obtained the explicit expression for} horizons (i.e. the Cauchy horizon $r_{-}$ and the event horizon $r_{+}=r_{h}$) {in the limit $\lambda \ll M$}. The bounds on the charge parameters $Q_{e}$ and $Q_{m}$, and PFDM parameter $\lambda$ for which a black hole exists is {also} obtained (see the shaded region in \cref{fig:QeQmlambda_region_plt}). It is observed that {the horizon of the black hole is very sensitive to PFDM parameter $\lambda$ as} the shaded region for which black hole exists (i.e the event horizon $r_{h}$), first shrinks {and then grows again as parameter $\lambda$ rises} 
{(see \cref{fig:QeQmlambda_region_plt,fig:hor_dm})}}. 

{We followed the Hamiltonian approach to study the dynamics of a charged moving around a charged dyonic black hole immersed in PFDM. To begin with, we first numerically analysed behaviour of $V_{eff}(r)$ for various combinations of black  hole parameters $Q_{e,m}$ and $\lambda$ in addition to the charge coupling parameters $\sigma_{e}$ and $g_{m}$. It is observed from the top right and bottom panel of \cref{fig:Veff_charged} that the maximum of the $V_{eff}(r)$ increases as well shifts towards the event horizon $r_{h}$ as an effect of parameters $Q_{e,m}$ and $\lambda$. Contrary to this, when parameters $Q_{e,m}$ and $\lambda$ are fixed and the coupling parameters $\sigma_{e}$ and $g_{m}$ vary the the maximum of $V_{eff}(r)$ reduces and shifts away from $r_{h}$ (see top left panel of \cref{fig:Veff_charged}). We also plotted the trajectories of charge particle in \cref{fig:traj}, it found that a charged particle have three different kind of orbits: bound, captured and escaping orbits, namely. On the other hand, a captured orbit occurs when the magnitude of magnetic charge $q_{m}$ increases, while a escaping orbit is observed for lager value of electric charge $q$. It is further observed that these orbits becomes bounded as consequence of increase PFDM parameter $\lambda$.
From \cref{fig:ang}, it is found that circular orbits of charge particles shift toward larger radii for as the parameter $g_{m}$ and $q_{e}$ increase.
 However, an opposite behavior is observed when PFDM parameter $\lambda$ increase. Finally, we presented the values ISCO parameters for the case of charged particles in \cref{tab:table1}, it is found that radius of ISCO $r_{ISCO}$ decreases with the rise in PFDM parameter $\lambda$ while it increases with the increase in the charge coupling parameter $g_{m}$ and $\sigma_{e}$. Whereas, the orbital angular momentum at ISCO $\mathcal{L}_{ISCO}$ decreases with the increase in parameter $\lambda$ keeping charge parameters $Q_{e,m}$ and coupling parameters $\sigma_{e}$ and $g_{m}$ of black hole fixed, an opposite behaviour is observed for $\mathcal{E}_{ISCO}$. Contrary to this, it is seen that the parameters $\mathcal{L}_{ISCO}$ and $\mathcal{E}_{ISCO}$ behave oppositely when increase the parameter $\sigma_{e}$ and $g_{m}$ keeping the black hole parameters $Q_{e,m}$ and $\lambda$ fixed. Additionally, it is found that the orbital velocity $v$ at ISCO always decreases as one move away as the value of the both the charge parameters $Q_{e,m}$ and PFDM $\lambda$ increases (see \cref{fig:vel}).}
 
 {To study the motion of massive spinning particle in the equatorial plane of the charged dyonic black hole immeresed PFDM, we used the Lagrangian approach over the MP approach for the reasons stated in the \cref{Sec:Appendix_A}. We explicitly obtained the expressions for the non-zero components of four-momentum $P^{\mu}$ and coordinated velocity $v^{\mu}$. For various combinations of parameters $Q_{e,m}, \lambda$, and $\mathcal{S}$, we numerically analysed the nature of effective Potential $V_{eff(+)}$ in detail (see \cref{fig:Veff1,fig:Veff2,fig:Veff3}) for both direct (i.e.  $\mathcal{L}=\mathcal{J}-\mathcal{S>0}$) and retrograde (i.e $\mathcal{L}=\mathcal{J}-\mathcal{S}<0$) trajectories. When comparing \cref{fig:Veff1,fig:Veff2}, it was discovered that $V_{eff(+)}$ is extremely sensitive to particle spin and the PFDM parameter $\lambda$, and showed  different behaviour when one parameter is fixed while the other is varied.}
 
 {Further, as Mathisson and Papapetrou \cref{eq:cov_der_four_mom,eq:cov_der_spin_tensor} lead to superluminal motion (i.e. space like behaviour) of the spinning particle  as they neglect the ``multi-pole" moments and consider only the ``spin-dipole" moment. In \cref{sec:ISCO_spinning}, we analysed the ISCOs parameters $r_{ISCO}, L_{ISCO}, E_{ISCO}$ and $\Omega_{ISCO}$ of spinning particle in detail to find the region where spinning particle motion is timelike and bring out the effect of PFDM $\lambda$ together with black hole charge parameters $Q_{e,m}$ on ISCOs. It is seen that the parameters $r_{ISCO}$ and $L_{ISCO}$ become smaller as the PFDM parameter $\lambda$ increases for the corresponding value of particle spin parameter $S$ and black hole charge parameters $Q_{e,m}$, which is consistent with the already established fact in \cite{Shaymatov21pdu,Narzilloev20a} that it can lead to an attractive force. Whereas, the parameters $E_{ISCO}$ and $L_{ISCO}$ increase (compare left column versus right column in \cref{fig:ISCO}). Analyzing more, the left column versus the right column of \cref{fig:ISCO}, it is interestingly observed that when the dark matter parameter $\lambda$ grows, the spinning particle enters the superluminal region (i.e. $U^{2}>0$, spacelike) for smaller values of spin, contrary to the fact established in the previous works that the superluminal region is reached for larger values of spin $\mathcal{S}$ if the dark matter field is absent \cite{Conde:2019juj}. 
 While when one moved down the columns in \cref{fig:ISCO}, the values of $r_{ISCO}, L_{ISCO}, E_{ISCO}$ and $\Omega_{ISCO}$ decrease with increase in the sum ($Q_{e}+Q_{m}$) of black hole charges. However, the limiting value of particle spin for which the motion is subluminal decreases slightly, contrary to what is observed when one moves along the row by varying the PFDM parameter $\lambda$ keeping $Q_{e,m}$ fixed.} 

{Finally, it is concluded that when the PFDM parameter $\lambda$ is very tiny (say 0.00001) and the black hole charges disappear  $Q_{e}=Q_{m}=0$, the ISCO parameters $r, L, E$, and $\Omega$ displayed in the top left panel of \cref{fig:ISCO} matched the Schwarzschild black hole scenario ( Fig. 2 of \cite{Conde:2019juj}), implying that for very small value of PFDM parameter $\lambda$ the charged dyonic black hole immersed in PFDM is identical to Schwarzschild black hole. It is also worth noting that our findings of a charged dyonic black hole immersed in PFDM reproduced the exact same conclusion as Jefremov \textit{et al.} \cite{Jefremov:2015gza} in the Schwarzschild limit.
On the other hand, in the case $Q_{m}=0$ and $\lambda=0.00001$, our results (see the middle panel on the left column of \cref{fig:ISCO}) coincided with Zhang \textit{et al.} \cite{Zhang:2017nhl} for Reissner-N\"{o}rdstrom black hole. These observations on the PFDM parameter $\lambda$ imply that for tiny enough values of $\lambda$ (say, $\lambda\leq 0.00001$), the charged dyonic black hole immersed in PFDM acts identically to its counterparts without PFDM. However, for large enough $\lambda$ values (say, $\lambda\geq 0.1$), the ISCO parameters $r_{ISCO}$ and $L_{ISCO}$ fall dramatically (as seen by comparing the left and right columns of \cref{fig:ISCO}), showing that small ISCOs are conceivable for large $\lambda$ values.}
 
{Thus, the work done here for the charged particles and spinning particles moving around the charged dyonic black hole submerged in PFDM is novel and intriguing since the PFDM effect on the spinning particle has not yet been documented in the literature.} Also, in the near future it  will become possible to detect and accurately measure gravitational waves emitted from extreme-mass ratio inspirals {as well as from  intermediate-mass-ratio inspirals} with the help of advanced space-based gravitational wave detectors like TianQin, Taiji, and LISA (Laser Interferometer Space Antenna). {As a consequence, with the advancement of technology, we will be able to acquire information about the ISCOs and our results, including the PFDM.} {As a result, the work done here on ISCOs will be important in better understanding the nature and initial condition of binary black hole mergers and their surrounding.}

\acknowledgments
{We are grateful to the referee for his insightful remarks and constructive suggestions that helped us to improve the clarity and precision of this paper.} 
S.S. acknowledges the support from Research Grant No. F-FA-2021-432 of the Uzbekistan Ministry for Innovative Development. P.S. acknowledges support under Council of Scientific and Industrial Research-RA scheme (Government of India).

\appendix
\section{A brief summary of Lagrangian theory of spinning particles in curved spacetime}
\label{Sec:Appendix_A}

The dynamics of a spinning particle in curved spacetime was first studied by Mathisson \cite{Mathisson:1937zz} and Papapetrou \cite{Papapetrou:1951pa} (hereinafter referred to as MP). Later, it as further extended by Tulczyjew \cite{Tulczyjew:1959a}, Taub \cite{Taub1948}, and Dixon \cite{Dixon:1964aa}. In the MP framework, the spinning test particle moving in curved spacetime follows a nongeodesic trajectory due the spin-orbit coupling. In their approach, the spinning particle is considered the particle of finite length that is much smaller than the characteristic length scale of spacetime; therefore, it assumed that the spinning possess a dipole moment in addition to the monopole moment that one needs to define the point particle.

In this appendix, for the sake of completeness (i.e., to make the paper self-contained), we present a brief summary of the Lagrangian theory of a spinning particle in curved spacetime.
For a detailed study of the theory of a spinning particle in curved spacetime, we encourage the reader to go through the articles \cite{Hanson:1974qy,Hojmanphdthesis:1975,Hojman:1976kn,Hojman:2012me}. In this work, we follow the Lagrangian approach to find the equations of motion for the spinning particles moving in the vicinity of a static spherically symmetric charged dyonic black hole immersed in PFDM. There are four main reasons for using the Lagrangian approach to find the equations of motion over the MP approach:
\begin{enumerate}[label=(\roman*)]
    \item Rather of adopting an \textit{adhoc or a posterior} definition, this approach allows for a precise description of the four-momentum. This is very useful, as the four-momentum and four-velocity are not parallel for the spinning particles moving in curved spacetime.
    \item The equations of motion are reparametrization covariant by nature.
    \item This theory give a natural interpretation of angular velocity, which helps us interpret the spin of a particle as a three-dimensional rotation.
    \item In this theory, the Casimir operators of the Poincar\'{e} group generated with the canonical momenta are constants of motion, allowing an easy description of the particle's mass $m$ and spin $S$.
\end{enumerate}

In this theory a Lagrangian $\mathcal{L}$ is constructed from four arbitrary invariant functions $\mathcal{C}_{1}\equiv u_{\mu}u^{\mu}$, $\mathcal{C}_{2}\equiv \sigma_{\mu\nu}\sigma^{\mu\nu}=-\text{tr}(\sigma^2)$, $\mathcal{C}_{3}\equiv u_{\alpha}\sigma^{\alpha\beta}\sigma_{\beta\nu}u^{\nu}$, and $\mathcal{C}_{4}\equiv \text{det}(\sigma)$,
\begin{align}
    \mathcal{L}\equiv\mathcal{L}\left(\mathcal{C}_{1},\mathcal{C}_{2},\mathcal{C}_{3},\mathcal{C}_{4}\right)=\sqrt{\mathcal{C}_{1}}L\left(\frac{\mathcal{C}_{2}}{\mathcal{C}_{1}},\frac{\mathcal{C}_{3}}{\mathcal{C}_{1}^{2}},\frac{\mathcal{C}_{4}}{\mathcal{C}_{1}^{2}}\right),
\end{align}
such that action $\mathcal{I}=\int \mathcal{L}d\tau$ is $\tau$-reparametrization invariant. Here, $u^{\mu}$ is the four-velocity $u^{\mu}\equiv dx^{\mu}/d\tau$ of the spinning particle, $\sigma^{\mu\nu}$ is the antisymmetric angular velocity tensor, $\tau$ is an affine parameter for the spinning particle position four-vector $x^{\mu}=x^{\mu}(\tau)$ and $L$ is an arbitrary function. It is worth noting here that, unlike the nonspinning particle case, it is not necessary for the spinning particle case to have a negative normalized four-velocity $\mathcal{C}_{1}$ in order to have a real $\mathcal{L}$ \cite{Hojman:2012me}.

The four-momentum $P_{\mu}$ and antisymmetric spin tensor $S_{\mu\nu}$ are defined via relations
\begin{align}
    P_{\mu}& \equiv \frac{\partial\mathcal{L}}{\partial u^{\mu}},\label{eq:Pmu}\\
    \text{and}\;\;\; S_{\mu\nu}&\equiv \frac{\partial\mathcal{L}}{\partial\sigma^{\mu\nu}}=-S_{\nu\mu}.\label{eq:Smunu}
\end{align}

As usual, by varying the action $\mathcal{I}$, the nongeodesic equations of motion for a spinning particle moving in curved spacetime are obtained \cite{Hojmanphdthesis:1975,Hojman:1976kn}
\begin{align}
    \frac{DP^{\mu}}{D\tau}&=-\frac{1}{2}R^{\mu}_{\nu\alpha\beta}u^{\nu}S^{\alpha\beta},\label{eq:cov_der_four_mom}\\
    \frac{DS^{\mu\nu}}{D\tau}&=S^{\mu\alpha}\sigma^{\nu}_{\alpha}-\sigma^{\mu\alpha}S^{\nu}_{\alpha}=P^{\mu}u^{\nu}-u^{\mu}P^{\nu},\label{eq:cov_der_spin_tensor}
\end{align}
where, $D/D\tau \equiv u^{\mu}\nabla_{\mu}$ and $R^{\mu}_{\nu\alpha\beta}$ are the covariant derivatives along $u^{\mu}$ and the Riemann tensor, respectively. Interestingly, these results hold for the arbitrary function $L$ as well and it is found that the dynamical variable four-momentum $P^{\mu}$ and spin tensor $S^{\mu\nu}$ are the generators of the Poincar\'{e} group. 

It is also shown in \cite{Hojman:2012me} that for a spinning particle moving in curved spacetime both its mass $m$ and spin $S$ are two conserved quantities defined as
\begin{align}
     m^{2}&\equiv -P_{\mu}P^{\mu},\label{eq:cons_P}\\
    S^{2}&\equiv \frac{1}{2}S_{\mu\nu}S^{\mu\nu}.\label{eq:cons_S}
\end{align}
One can see here that conservation of $S^{2}$ is coming solely from the \cref{eq:cov_der_spin_tensor} equation of motion by contracting  \cref{eq:cov_der_spin_tensor} with the covariant component of spin tensor $S_{\mu\nu}$. It is worth to emphasizing here that conservation of particle spin comes naturally in the Lagrangian theory \cite{Hojmanphdthesis:1975}, in contrast with the extended MP formulation \cite{Dixon:1964aa}, which requires an extra assumption.

Furthermore, simply glancing at \cref{eq:cov_der_four_mom,eq:cov_der_spin_tensor} reveals that there are more unknown variables than there are equations. Toover come this difficulty, the TSSC \cite{Tulczyjew:1959a}
\begin{align}
    P_{\mu}S^{\mu\nu}=0 \label{eq:Tulc_SSC}
\end{align}
is used. The above \cref{eq:Tulc_SSC} helps us set three of the six independent components of spin tensor to zero (i.e. $S^{0i}=0$) for a particular frame of reference. Hence, it gives the freedom to fix the worldline  describing the path of a spinning particle.  The components of $S^{0i}$ are associated with the mass dipole moment of the spinning particle, so setting these equal to zero in some particular frame of reference fixes the center of mass of the spinning particle in that frame of reference (for a detailed discussion on SSC, we request reader to see \cite{Costa:2014nta} and the references therein). Also, it is shown in \cite{Schattner:1979vn} that \cref{eq:Tulc_SSC} defines a unique worldline of the spinning particle in a curved spacetime. In addition to the four reasons stated earlier for using the Lagrangian theory to find equations of motion, it worth noting here that if function $L$ is chosen wisely, one would derive the TSSC from the Lagrangian \cite{Hojman:2012me}, which is another motivation to use the Lagrangian theory approach.
{Now, one can define a normalized four-momentum as
\begin{align}
V^{\mu}=\frac{P^{\mu}}{m},\label{eq:normalized_4momentum}
\end{align}
 such that it satisfies the conservation relation ($V_{\mu}V^{\mu}$=-1). As the four-velocity $u^{\mu}$ is not the conserved quantity for the case of spinning particle. Hence, one need to bridle the $u^{\mu}$ as
 \begin{align}
     u_{\mu}u^{\mu}<0,
 \end{align}
 in addition to SSC \cref{eq:Tulc_SSC} so that the spinning particle will have a timelike motion. For convince, one may choose
 \begin{align}
     u_{\mu}V^{\mu}=-1,\label{eq:rel_bw_u_n_V_1}
 \end{align}
as pointed in \cite{Saijo:1998mn}.
Now, using \crefrange{eq:cov_der_four_mom}{eq:rel_bw_u_n_V_1}, a relation between a $u^{\mu}$ and momentum $V^{\mu}$ can be establish which reads as
\begin{align}
    u^{\mu}-V^{\mu}=\frac{2S^{\mu\nu}R_{\nu\alpha\beta\gamma}V^{\alpha}S^{\beta\gamma}}{4 m^{2}+R_{\delta\rho\sigma\kappa}S^{\delta\rho}S^{\sigma\kappa}}.\label{eq:rel_bw_u_n_V_2}
\end{align}}

Additionally, more conserved quantities can be found depending on the geometry of the spacetime by using
\begin{align}
    \mathfrak{C}_{\xi}\equiv P^{\mu}\xi_{\mu}-\frac{S^{\mu\nu}\xi_{\mu;\nu}}{2},\label{eq:Killing}
\end{align}
where, $\xi_{\mu}$ is a Killing vector that satisfies the relation
\begin{align}
    2\xi_{\left(\mu;\nu\right)}=0.
\end{align}

In order to make this paper self-contained, we presented a brief overview of Lagrangian theory. The equations of motion of the spinning particle moving in the vicinity of a charged dyonic black hole immersed in PFDM are determined using \crefrange{eq:Pmu}{eq:Killing}.

\section{Explicit expressions of ISCO equations and superluminal constraint}
\label{Sec:Appendix_B}
Here, we present the explicit form of ISCO \cref{eq:ISCO1,eq:ISCO2,eq:ISCO3} as a function of the radial coordinate $r$, spin per unit mass $\mathcal{S}$, total angular per unit mass $\mathcal{J}$, energy per unit mass $\mathcal{E}$, and black hole parameters $M, Q_{e}$ and $Q_{m}$:
\vspace{-1.0cm}
\begin{widetext}
\begin{align}
  W_{eff}&=\frac{1}{\left(2 r^5+r \mathcal{S}^2 \mathcal{Y}\right)^2}\Bigg[\left(2 \mathcal{E} r^5+\mathcal{J} r \mathcal{S Y}\right)^2 +\left(4 r^6 (\mathcal{J-E S})^2+\left(2 r^4+\mathcal{S}^2 \mathcal{Y}\right)^2\right) \mathcal{X}\Bigg]=0,\label{eq:ISCO11}\\
  \frac{dW_{eff}}{dr}&=\frac{1}{\left(2 r^5+r \mathcal{S}^2 \mathcal{Y}\right)^3} \Bigg[-8 r^6 \mathcal{X} (\mathcal{J}-\mathcal{E} \mathcal{S})^2 \left(2 r^4-\mathcal{S}^2 \left(2 \lambda  r \ln{ \left(\frac{r}{\left| \lambda \right| }\right)}-4 M r+6 Q_{e+m}^2-3 \lambda  r\right)\right)\nonumber
  \\
  &-2 r^2 \mathcal{S}^2 \eta \Psi^2+2 \mathcal{J} r^2 \mathcal{S} \Psi\eta \left(\lambda  r \mathcal{S}^2 \ln{\left(\frac{r}{\left| \lambda \right| }\right)}-r \mathcal{S}^2 (\lambda +2 M)+2 \mathcal{S}^2 Q_{e+m}^2+2 r^4\right)\nonumber
  \\
  &\;+\mathcal{Y} \left(2 r^4+\mathcal{S}^2 \mathcal{Y}\right) \left(4 r^6 (\mathcal{J}-\mathcal{ES})^2+\left(2 r^4+\mathcal{S}^2 \mathcal{Y}\right)^2\right)\Bigg]=0,\label{eq:ISCO21}\\
  \frac{d^2 W_{eff}}{dr^2}&=-\frac{1}{\left(2 r^5+r \mathcal{S}^2 \mathcal{Y}\right)^4}\Bigg[-8 r^6 \mathcal{X} (\mathcal{J}-\mathcal{E S})^2 \left(\lambda  r \mathcal{S}^2 \ln{ \left(\frac{r}{\left| \lambda \right| }\right)} \left(-r \mathcal{S}^2 (19 \lambda +24 M)+36 \mathcal{S}^2 Q_{e+m}^2-36 r^4\right)\right.\nonumber
  \\
  &+6 \lambda ^2 r^2 \mathcal{S}^4 \ln ^2{\left(\frac{r}{\left| \lambda \right| }\right)}+2 \bigg(r^2 \mathcal{S}^4 \left(8 \lambda ^2+12 M^2+19 \lambda  M\right)-r \mathcal{S}^4 (31 \lambda +36 M) Q_{e+m}^2+r^5 \mathcal{S}^2 (29 \lambda +36 M)\nonumber
  \\
  &-60 r^4 \mathcal{S}^2 Q_{e+m}^2+30 \mathcal{S}^4 \left(Q_{e+m}^2\right)^2+6 r^8\bigg)\bigg)+16 r^6 \mathcal{Y} (\mathcal{J}-\mathcal{E} \mathcal{S})^2 \left(2 r^4+\mathcal{S}^2 \mathcal{Y}\right) \bigg(2 r^4-\mathcal{S}^2 \bigg(2 \lambda  r \ln{\left(\frac{r}{\left| \lambda \right| }\right)}\nonumber
  \\
  &-4 M r+6 Q_{e+m}^2-3 \lambda  r\bigg)\bigg)+\left(2 r^4+\mathcal{S}^2 \mathcal{Y}\right)^2 \left(4 r^6 (\mathcal{J}-\mathcal{E}\mathcal{S})^2+\left(2 r^4+\mathcal{S}^2 \mathcal{Y}\right)^2\right) \bigg(2 \lambda  r \ln{\left(\frac{r}{\left| \lambda \right| }\right)}-4 M r\nonumber
  \\
  &+6 Q_{e+m}^2-3 \lambda  r\bigg)-2 \mathcal{J S} \left(2 r^5+r \mathcal{S}^2 \mathcal{Y}\right)^2 \left(\left(12 \lambda  r \ln{\left(\frac{r}{\left| \lambda \right| }\right)}-24 M r+40 Q_{e+m}^2-19 \lambda  r\right) \bigg(\mathcal{J} \lambda  r \mathcal{S} \ln{\left(\frac{r}{\left| \lambda \right| }\right)}\right.
  \nonumber
  \\
  &\left.+2 \mathcal{E} r^4+\mathcal{J S} \left(-r (\lambda +2 M)+2 Q_{e+m}^2\right)\bigg)+\mathcal{J S} \left(3 \lambda  r \ln{\left(\frac{r}{\left| \lambda \right| }\right)}-6 M r+8 Q_{e+m}^2-4 \lambda  r\right)^2\right)\nonumber
  \\
  &-2 \left(2 \mathcal{E} r^5+\mathcal{J} r \mathcal{S} \mathcal{Y}\right)^2 \left(\mathcal{S}^2 \left(-12 \lambda  r \ln{\left(\frac{r}{\left| \lambda \right| }\right)}+24 M r-40 Q_{e+m}^2+19 \lambda  r\right)\left(\lambda  r \mathcal{S}^2 \ln{\left(\frac{r}{\left| \lambda \right| }\right)}-r \mathcal{S}^2 (\lambda +2 M)\right.\right.\nonumber
  \\
  &+2 \mathcal{S}^2 Q_{e+m}^2+2 r^4\bigg)+3 \mathcal{S}^4 \left(3 \lambda  r \log \left(\frac{r}{\left| \lambda \right| }\right)-6 M r+8 Q_{e+m}^2-4 \lambda  r\right)^2\bigg)+8 \mathcal{J} r^2 \mathcal{S}^3 \bigg(3 \lambda  r \ln{\left(\frac{r}{\left| \lambda \right| }\right)}-6 M r\nonumber
  \\
  &+8 Q_{e+m}^2-4 \lambda  r\bigg)^2 \left(\lambda  r \mathcal{S}^2 \ln{\left(\frac{r}{\left| \lambda \right| }\right)}-r \mathcal{S}^2 (\lambda +2 M)+2 \mathcal{S}^2 Q_{e+m}^2+2 r^4\right)\bigg(\mathcal{J} \lambda  r \mathcal{S} \ln{\left(\frac{r}{\left| \lambda \right| }\right)}+2 \mathcal{E} r^4\nonumber
  \\
  &+\mathcal{J S} \left(-r (\lambda +2 M)+2 Q_{e+m}^2\right)\bigg)\Bigg]=0,\label{eq:ISCO31}
\end{align}
\end{widetext}
\begin{widetext}
and the explicit form of $U^{2}$ \cref{eq:U2} using \cref{eq:F,eq:dotr1,eq:dotphi} reads
\begin{align}\label{eq:U2_1}
  U^{2}&= -F(r)+\frac{4 r^2 \mathcal{X}^2 (J-\mathcal{E S})^2 \left(-2 \lambda  r \mathcal{S}^2 \log \left(\frac{r}{\left| \lambda \right| }\right)+r \mathcal{S}^2 (3 \lambda +4 M)-6 \mathcal{S}^2 Q_{e+m}^2+2 r^4\right)^2}{\left(\lambda  r \mathcal{S}^2 \log \left(\frac{r}{\left| \lambda \right| }\right)-r \mathcal{S}^2 (\lambda +2 M)+2 \mathcal{S}^2 Q_{e+m}^2 +2r^4\right)^2 \Psi^2}\nonumber
  \\
  &-\Bigg[\frac{\mathcal{X} \left(\left(2 \mathcal{E} r^5+\mathcal{J} r \mathcal{S} \mathcal{Y}\right)^2+\mathcal{X} \left(4 r^6 (\mathcal{J}-\mathcal{E S})^2+\left(2 r^4+\mathcal{S}^2 \mathcal{Y}\right)^2\right)\right) }{r^2 \left(2 r^5+r \mathcal{S}^2 \mathcal{Y}\right)^2\Psi^2}\Bigg]\nonumber
  \\
  &\times \left(\lambda  r \mathcal{S}^2 \ln{\left(\frac{r}{\left| \lambda \right| }\right)}-r \mathcal{S}^2 (\lambda +2 M)+2 \mathcal{S}^2 Q_{e+m}^2+2 r^4\right)^2,
 \end{align}
where,
\begin{align}
   \eta &=-3 \lambda  r \ln{\left(\frac{r}{\left| \lambda \right| }\right)}+6 M r-8 Q_{e+m}^2+4 \lambda  r,\\
   \Psi&=\mathcal{J} \lambda  r \mathcal{S} \ln{\left(\frac{r}{\left| \lambda \right| }\right)}+2 \mathcal{E} r^4+\mathcal{J S} \left(-r (\lambda +2 M)+2Q_{e+m}^2\right).
\end{align}
\end{widetext}

\bibliographystyle{apsrev4-1}  
\bibliography{gravreferences,Ref,Spinning}
\end{document}